\documentclass[]{aa}  
\usepackage{graphicx,natbib}
\usepackage{epsfig,epstopdf}
\usepackage{ulem}
\usepackage[]{inputenc} 
\usepackage[T1]{fontenc}
\bibpunct{(}{)}{;}{a}{}{,}
\usepackage{txfonts}
\include{def}
\usepackage{changebar}
\usepackage{booktabs}
\usepackage{caption}
\usepackage{ctable}
\usepackage{mathrsfs}
\usepackage{ctable}
\usepackage{multirow}
\usepackage{tablefootnote}

\newcommand{\ME}{M$_{\oplus}$} 
\newcommand{\Mearth}{\text{M}_{\oplus}} 

\newcommand{\rhocgs}{g/cm$^3$}

\newcommand{\Mcore}{M_{\text{core}}}
\newcommand{\Menv}{M_{\text{env}}}

\newcommand{\MHHe}{M_{\text{HHe}}}
\newcommand{\Mthresh}{M_{\text{thresh}}}

\newcommand{\Mzenv}{M_{\text{Z,env}}}
\newcommand{\Mzdot}{\dot{M}_{\text{Z}}}
\newcommand{\Mdotcore}{\dot{M}_{\text{core}}}
\newcommand{\Rcore}{R_{\text{core}}}
\newcommand{\Rp}{R_{\text{P}}}
\newcommand{\Mp}{M_{\text{P}}}

\newcommand{\Ztot}{Z_{\text{tot}} }

\newcommand{\Zenv}{Z_{\text{env}} }
\newcommand{\Zsun}{Z$_{\odot}$}

\newcommand{\Lcore}{L_{\text{core}}}

\newcommand{\Lgas}{L_{\text{gas}}}

\newcommand{\tcross}{t_{\text{cross}}}
\newcommand{\tauhhe}{\tau_{\text{HHe}}} 
\newcommand{\MdotHHe}{\dot{M}_{\text{HHe}}}
\newcommand{\Mcross}{M_{\text{cross}}}

\newcommand{\trun}{t_{\text{run}}}

\begin{document}

   \title{Planet formation with envelope enrichment: new insights on planetary diversity}

    \author{Julia Venturini\thanks{Currently at the Institute for Computational Science, University of Zurich, Winterthurerstrasse 190, 8057 Zurich, Switzerland}  \inst{} 
       \and
       Yann Alibert \inst{}
       \and
         Willy Benz \inst{}
     }
\institute{Physikalisches Institut, Universit\"at Bern, Sidlerstrasse 5, 3012 Bern, Switzerland \\ \email{julia@physik.uzh.ch}
}


   \date{\today}

 
  \abstract{}
   {We compute, for the first time, self-consistent models of planet growth including the effect of envelope enrichment. The change of envelope metallicity is assumed to be the result of planetesimal disruption or icy pebble sublimation.}
   {We solve internal structure equations taking into account global energy conservation for the envelope to compute in-situ planetary growth. We consider different opacities and equations of state suited for a wide range of metallicities.}
   {We find that envelope enrichment speeds up the formation of gas giants. It also explains naturally the formation of low and intermediate mass objects with large fractions of H-He ($\sim$ 20 - 30 \% in mass). High opacity models explain well the metallicity of the giant planets of the solar system, whereas low opacity models are suited for forming small mass objects with thick H-He envelopes and gas giants with sub-solar envelope metallicities. We find good agreement between our models and the estimated water abundance for WASP-43b. For HD 189733b, HD 209458b and WASP-12b we predict fractions of water larger than what is estimated from observations, by at least a factor $\sim 2$.}
   {Envelope enrichment by icy planetesimals is the natural scenario to explain the formation of a large variety of objects, ranging from mini-Neptunes, to gas giants. We predict that the total and envelope metallicity decrease with planetary mass.}


\titlerunning{Planet formation with envelope enrichment}
\authorrunning{Venturini et al.}
   \maketitle

\section{Introduction}\label{intro}
The core accretion model \citep{PerriCam74, mizuno80, BP86, P96} is the most accepted scenario to explain the formation of a vast diversity of planets \citep[e.g,][]{Alibert05, Mord09, Guilera11}. The central idea of the core accretion model can be summarised as follows. First, a solid core must be formed from the accretion of planetesimals/pebbles. Once this core reaches approximately a lunar mass, the core gravity is strong enough to start to bind some gas from the protoplanetary disk. Thus, from this stage on, the protoplanet keeps growing by accreting both solids (planetesimals/pebbles) and gas (basically H-He). Planet formation models typically assume, for simplification, that solids and gas do not mix: all the solids deposit their mass and energy at the top of the core, and the primordial H-He is collected above, building the atmosphere (or envelope). This is, of course, a very strong and unrealistic simplification: bolides that traverse a gaseous atmosphere undergo thermal ablation and mechanical breakup. Hence, volatile material can vaporise and mix with the primordial H-He, changing the composition of the envelope during the formation of a planet.

If planetesimal/pebble disruption did not occur during formation, then the envelope metallicities of planets should be rather sub-stellar, because the gas accreted into the planets should in principle be metal-poor compared to the central star (the metals condense to form planetesimals/pebbles). This is not what is observed in the solar system, where the giant plants show some level of envelope enrichment \citep{Irwin14, galileo_probe, Guillot14}. The alternative hypothesis to planetesimal/pebble dissolution for explaining envelopes enriched in heavy elements, is core erosion \citep{Wilson12}. From an energetic point of view, it is possible to mix part of the core upwards \citep{Guillot04}. In addition, core material is miscible in metallic hydrogen \citep{Wilson12}, which allows for the heavy elements (if they can be lifted up) not to settle to re-form a core. Regarding the mixing of an initial core within the gaseous envelope, \citet{Vazan15} showed that this process is favoured if an initial compositional gradient exists in the interior of the planet. Thus, the mixing of heavy elements on the planetary envelope seems to be more likely if the heavy elements are not initially concentrated in well-defined core. The formation of such a diffuse core requires planetesimal dissolution in the deep envelope during the formation of the planet. Hence, even if core erosion could play a relevant role in mixing heavy elements in the planetary envelope, this process seems to demand as well that the envelope is initially enriched by planetesimal/pebble dissolution.

The problem of considering an envelope that is enriched with respect to stellar values during the formation of a giant planet has been raised since the very early studies of planet formation. Already in 1986, \citet{BP86} mentioned, among other two problems that remained to be solved ``the fact that the molecular weight of the envelope is expected to increase with time as some of the icy planetesimals dissolve in it", and added that this problem ``could significantly change the accretion scenario". Indeed, \citet{Stevenson82} showed that the critical core mass (the mass required to trigger rapid accretion of gas) is reduced when the envelope mean molecular weight increases. Moreover, \citet{Wuchterl93} showed a direct dependence of the critical core mass with the adiabatic gradient. The adiabatic gradient is expected to decrease when chemical reactions take place. This necessarily occurs when abundant elements such as H, C, O are present in the envelope. Therefore, the effect of polluting the primordial envelopes reduces the critical core mass  not only due to the increase of mean molecular weight but also via a reduction of the adiabatic gradient \citep{HI11}. Another effect that can reduce the adiabatic gradient is the condensation of species. We showed \citep{Venturini15} that if a planet forms in cold regions of the disk, such that the temperatures and pressures are low enough for certain species to condense in the atmosphere of the protoplanet, then this effect leads to an even larger reduction of the critical core mass. 

Despite its expected importance for the formation of giant planets, the effect of envelope enrichment has so far never been implemented in a self-consistent way in any evolutionary calculation of planet formation. This is of course a consequence of the difficulty in modelling all the processes involved, which include: planetesimal thermal ablation and dynamical breakup, mass and energy deposition in different envelope layers, and the inclusion of a self-consistent change in the envelope's microphysics (opacities and equation of state) as the envelope metallicity evolves. 

The magnitude of the enrichment depends on the accretion rate of solids, and on their size and strength properties. The smaller and more porous the bolide, the easier it is to disrupt it and mix it with the envelope gas when crossing the atmosphere \citep{Podolak88}. This tells us that the effect of envelope enrichment is more relevant for smaller bolides. Hence, when growing planets from cm-m size particles \citep[the so-called \textit{pebbles},][]{Lambrechts12, Lambrechts14}, including this effect is necessary.

In this work we compute, for the first time, the in-situ growth of a planet taking into account envelope enrichment by icy planetesimals/pebbles. Given the uncertainties in the initial size distribution of planetesimals, we test the effect of envelope enrichment corresponding to a wide range of particle sizes. We solve internal structure equations using global energy conservation arguments, and taking into account the changes of envelope metallicity in the opacities and equation of state. In Sect. \ref{numerics} we explain the numerical method, Sect. \ref{assumptions} is devoted to discussing our assumptions, in Sect. \ref{resultados} and \ref{P96_section} we show and analyse our results, in Sect. \ref{implications} and \ref{predictions} we discuss our main results and predictions. We summarise our conclusions in Sect. \ref{conclusions}.

\section{Methodology}\label{numerics}
We use the numerical method developed in \citet{Venturini15} to solve the equations of hydrostatic equilibrium, mass conservation and heat transport. We apply the usual Schwarzschild criterion to distinguish between radiative and convective layers, adopting an adiabatic gradient for the latter case. 

Regarding energy conservation, we use global energy conservation arguments \citep{Mordasini12b, Fortier13, Piso14} to find the total luminosity radiated away by the protoplanet. In the following subsection we show how we compute this total luminosity. 

\subsection{Computation of the total luminosity}
\begin{figure}
\begin{center}
\includegraphics[width=1.01\columnwidth]{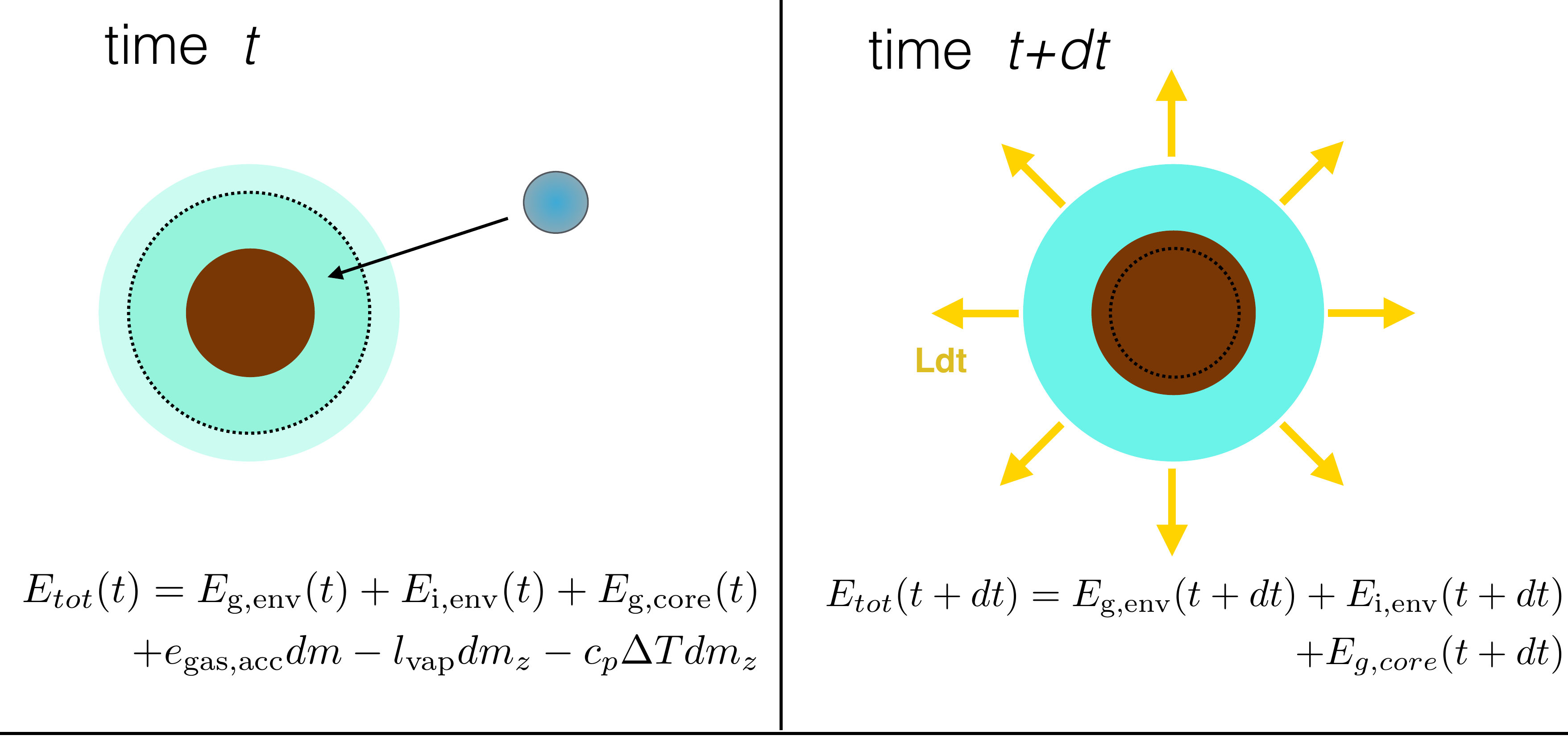} 
   \caption{Sketch representing the total energy of the system at time $t$ and $t+dt$. See main text for the explanation of the different terms.}
    \label{sketch_lum}
\end{center}
\end{figure}

To explain how we compute the total luminosity ($L$), let us analyse the total energy of the system at time $t$ and at time $t+dt$, as it is sketched in Fig. \ref{sketch_lum}. 

At a given time $t$, the total energy of the system is given by the sum of \footnote{The planetesimals are assumed to have started with zero velocity at infinity, therefore their total energy is zero. The release of their gravitational potential energy, which heats the envelope, appears as the term of change of gravitational potential energy of the core, as we show later in the main text.}:
\begin{itemize}
\item  the total energy of the envelope (gravitational plus internal):   $E_{\text{g,env}}(t) + E_{\text{i,env}}(t)$
\item  the gravitational potential energy of the core (we neglect heat sources coming from the core \footnote{The change of $E_{\text{g,core}}(t)$ with time has typically, minimum values of $10^{26}$ erg/s, whereas the change of $E_{\text{i,core}}(t)$ with time reaches values of $10^{23}$ erg/s \citep{Lopez14} and is therefore negligible. The cooling of the core should, on the other hand, be considered once the heat due to planetesimal bombardment ($dE_{\text{g,core}}/dt$) ceases, as is the case, for example, in evolutionary simulations.}): $E_{\text{g,core}}(t)$
\item the total energy of the gas layer that will be accreted, represented by the term  $e_{\text{gas,acc}} dm$, where $e_{\text{gas,acc}}$  is the total energy of the layer per unit mass, and $dm$ the mass of gas that will be accreted
\item the energy that needs to be subtracted from the envelope in order to vaporise and thermalise the ices of the planetesimals that will be mixed in the envelope. The vaporisation is given by the term $-l_{\text{vap}} dm_z$, $l_{\text{vap}}$ being the latent heat of vaporisation of the ice per unit mass, and $dm_z$ the mass of ice remaining in the envelope \footnote{We assume the ices to be water ice. We explain better this assumption in Sect. \ref{Beta} and Sect. \ref{EOS}.}. The thermalisation term is given by $c_p \Delta T dm_z $, being $c_p$ the specific heat capacity of the ice and $\Delta T$ the change of temperature from 0$^{\circ}$C to the mean envelope temperature.
\end{itemize}
At time $t + dt$, all the mass of the planetesimals and gas that were accreted in the time elapsed ($dt$) are now part of the protoplanet, so the total energy is just the gravitational potential of the core and envelope, plus the internal energy of the envelope.

Our protoplanet is not a closed system, it is embedded in a protoplanetary disk, which pushes the outermost layer of the protoplanet, making work represented by $-P_{\text{disk}} dV$. Also, the protoplanet cools, radiating energy away by photons ($- Ldt$). Hence,

 \begin{equation}
 E_{tot} (t+dt) - E_{tot} (t) = - L dt - P_{\text{disk}} dV.
 \end{equation}
 
Regrouping  terms, we find 
\begin{align}
L = &- \frac{dE_{\text{g,core}}}{dt} -\frac{dE_{\text{env}}}{dt} + e_{\text{gas,acc}} \dot{M}_{\text{gas}} - P_{\text{disk}} \frac{dV}{dt} \nonumber \\ 
& - (1-\beta) (l_{\text{vap}} + c_p \Delta T) \dot{M}_{z}. 
\end{align}

The first term is just the change of gravitational potential energy of the core, the second is the change of total energy of the envelope, the next two are the surface terms already explained, and the last are the ``ice heating" terms, also explained above. We call $\beta$ the mass fraction of planetesimals that sinks to the core, that is why a factor $1 - \beta$ (what remains in the envelope) is needed in front of the ``ice heating" terms. 

Now let us further develop the first two terms, which are always at least two orders of magnitude larger than the others.
To compute the first term (which we will call hereafter $\Lcore$), we make use of the fact that the gravitational potential energy of a sphere of uniform density is: 

\begin{equation}
E_{\text{core}} = - \frac{3}{5} \frac{G \Mcore^2}{\Rcore}.
\end{equation}

Differentiating with respect to time and using the fact that the core density is constant\footnote{We assume, throughout this work, a constant density for the core, an usual practice in most planet formation works. If the core were compressible, an extra term in $\Lcore$ should be included, reflecting the change of core gravitational energy associated with the change of density. Nevertheless, since the core is in solid phase, that term is not really relevant.}, one can readily find:

\begin{equation}
\Lcore = - \frac{dE_{\text{g,core}}}{dt} = \frac{G \Mcore \Mdotcore}{\Rcore}.
\end{equation}

If we call $\Mzdot$ the total accretion rate of solids, since $\beta$ is the mass fraction of planetesimals/pebbles that sinks to the core, then $\Mdotcore = \beta \Mzdot$.

We express the second term as $\Lgas$: 
\begin{equation}
\Lgas = - \frac{E_{\text{env}}(t+dt) - E_{\text{env}}(t)}{dt} .
\end{equation}

The problem with the computation of $\Lgas$ is that the energy of the envelope at time $t+dt$ is not known before obtaining the structure at $t+dt$. Therefore, we must make a \textit{guess} of  $E_{\text{env}}(t+dt)$ to be able to have $L(t+dt)$ for computing the structure at $t+dt$. This guess is performed following  \citet{Mordasini12b} and \citet{Fortier13}. These authors assumed a similar functional form for the total energy of the envelope as the one of the gravitational potential energy of the core of uniform density. Hence,

\begin{equation}\label{kenv}
E_{\text{env}} = - k_{\text{env}} \Menv g,
\end{equation}
where $g$ is a mean gravity, defined by $g =  \frac{1}{2} G(\Mcore / \Rcore + M_{\text{P}} / R_{\text{P}} )$.

For a given converged structure at time $t$, we know the total energy of the envelope at this time. Hence, Eq.(\ref{kenv}) defines  $k_{\text{env}}$ at time $t$, which is then used for the total energy of the envelope at time $t+dt$. This assumption is quite good during most of the growth of the planet, since $k_{\text{env}}$ does not practically vary from one timestep to another. Indeed, to test this we have run simulations using a predictor-corrector scheme: we first compute the structure of the envelope using the $k_{\text{env}}$ from the previous timestep, $k_{\text{env}}$(t - dt), then compute the resulting $k_{\text{env}}$ of the actual timestep,  $k_{\text{env, corr}}$, and finally recompute the structure of the planet at time $t$ using $k_{\text{env}}$  = 0.5 ($k_{\text{env}}$ (t - dt)+ $ k_{\text{env, corr}})$.

\subsection{Envelope enrichment and iteration scheme}\label{iterZ}
For the initial conditions, we assume that the gas in the disc - and therefore the planetary envelope- is made of hydrogen and helium. This is justified by the fact that we are forming planets beyond the iceline, and thus, most of the metals have already condensed into solids\footnote{Strictly speaking, volatile species with condensation temperatures below the one of water (e.g, CO) could exist in the gas phase beyond the iceline. Nevertheless, the mass contribution of these species to the total gaseous disc is expected to be negligible \citep{Amaury15}.}, which exist in the disc in the form of embryos, planetesimals or pebbles.
 
We start all simulations with $\Mcore$ = 0.01 $\Mearth$. The core grows at a given accretion rate $\Mzdot$ (see Sect. \ref{AccretionRates} for the different schemes adopted to compute $\Mzdot$). When the core reaches a threshold value of $\Mthresh$, we assume that the core keeps growing by an amount $\Delta \Mcore = \beta \Mzdot dt $ while the amount $\Delta \Mzenv$ = $(1-\beta) \Mzdot dt $ remains uniformly mixed in the envelope. The dependence of $\Mthresh$ on the planetesimals' properties is discussed in Sect. \ref{choice_mthresh}.

At the beginning of each timestep, an initial guess of envelope metallicity is assumed in order to compute the corresponding equilibrium envelope mass (the envelope mass depends on the core mass and radius, total luminosity and envelope metallicity). For the computation of $\Menv$ we use the second numerical scheme described in \citet{Venturini15}: iteration on $\Menv$ for a given core mass and radius, and $\Zenv$. The total luminosity is recalculated at each iteration on $\Menv$, because $\Lgas$ depends on $\Menv$, as explained in the above subsection. 
The envelope metallicity is taken as uniform throughout the envelope and is defined as:
\begin{equation}\label{Zenv}
\Zenv = \frac{\Mzenv}{\Menv} ,
\end{equation}
$\Mzenv$ being the mass of volatiles that remains mixed in the envelope.

Once convergence in $\Menv$ is achieved, the envelope metallicity is updated to the new $\Menv$ found. Afterwards, an extra iteration on $\Zenv$ is required in order to have the metallicity self-consistently defined (and thus, mass conservation of metals guaranteed).

\section{Assumptions}\label{assumptions}
\subsection{Choice of $\beta$}\label{Beta}
We recall that $\beta$ is the mass fraction of the incoming planetesimals that is assumed to reach the core (hence, the fraction $1-\beta$ remains homogeneously mixed in the envelope). For our nominal model, we take $\beta = 0.5$. This choice is based on the refractory/volatile content of comets \citep{Lambrechts14, Amaury15}. Of the material falling in, we assume that only ices mix in the envelope and that the refractory sinks into the core. We assume for simplification that the ices are just water ice, because it is the main volatile species in comets \citep{Kofman15}, and because of self-consistency with our equation of state (see Sect. \ref{EOS}). The choice of $\beta = 0.5$ is a strong assumption, but we expect the ices to be the first component of the incoming planetesimals to be vaporised in the envelope. Moreover, it has been shown that water and hydrogen remain well mixed in the interior of giant planets, for pressures larger that 1 GPa and temperatures larger than 2000 K \citep{Soubiran15}. So at least for the mentioned ranges, the assumption of water being homogeneously mixed in the envelopes is a fair one. However, these ranges do not necessary cover all the pressure-temperature values existent during formation \citep[see Fig. 5 of][]{Venturini15}, which can extend from $P\sim 10^{-2}$ Pa to P $\sim$10 GPa and from T$ \sim$ 100 K to T$ \sim$ 10,000 K. The miscibility of water into hydrogen for low values of pressure and temperature should be tested in the future.

\subsection{Envelope microphysics}\label{microfisca_paper2}
\subsubsection{Opacities}\label{opac_paper2}
In \citet{Venturini15} we assumed a total opacity resulting from contributions from dust and gas. We used dust opacities from \citet{Semenov03} and a gas opacity from tables computed by Ferguson for arbitrary metallicity, based on the original calculations of \citet{AlexFerg94}. The dust opacities from \citet{Semenov03} are suited for protoplanetary disks, and therefore do not consider grain growth and settling in a planetary envelope. This is taken into account in the analytical opacities calculated by \citet{Mordasini14}, which we adopt in this work. The dust opacities of \citet{Mordasini14} have the additional advantage of being independent on the accretion rate of metals (an equilibrium among the deposition of grains, grain growth and settling was found in those calculations), making these opacities independent  of the envelope metallicity and therefore, suited for any value of $\Zenv$.

Regarding gas opacities, new calculations by \citet{Freedman14} are suited for planetary envelopes and for a wide range of metallicities ( $0 \lesssim \Zenv\lesssim 0.5$). Compared to the opacities computed by Ferguson, they have the advantage of not having to be extrapolated for T$\le$1000 K, and also the fact that an analytical fit is available, which is very useful to reduce computational time. The domain of validity in pressure and temperature of these opacities is, respectively,  1 $\mu$bar to 300 bars and 75 K to 4000 K. These ranges do not reach the high temperatures and pressures that are typical at the surface of the core, but in practice, in that domain of high pressure and temperature the envelopes are always convective. Therefore, even taking a constant extrapolation in these regions does not affect the internal structure computation.

In any case, it is well known that opacities affect drastically the timescale to form a giant planet \citep{P96,Hubi05,Ikoma00}. This is why we run comparison tests using different extreme combinations of dust and gas opacities. On one side, we take the old combination of Semenov + Ferguson, and on the other, Mordasini + Freedman. The results concerning the crossover mass are summarised in Table \ref{opacHHe} for a solar composition envelope and a constant accretion rate of metals of $10^{-6}$ $\Mearth$/yr . The notorious decrease in the crossover mass when using the new opacities is mainly due to the fact that the dust opacities from \citet{Mordasini14} are much lower than the ones from \citet{Semenov03} (whose values are very similar to the interstellar ones). We repeat this comparison test taking into account envelope enrichment in Sect. \ref{BLFerg_Zenv}. For all results shown before, we adopt the new opacities of Mordasini for the dust and Freedman for the gas.

\begin{table}
\centering
\begin{tabular}{  l   c }
\hline\hline\
   Opacities  &  Crossover Mass [\ME]\\
  \hline
   Semenov + Ferguson &  17 \\
   Mordasini + Freedman & 6.9  \\
\hline  
\end{tabular}
\caption{Crossover mass for Z= \Zsun, different opacities and $\Mzdot$ = $10^{-6}$ \ME/yr.}
\label{opacHHe}
\end{table} 

 \subsubsection{Equation of State (EOS)}\label{EOS}
We assume in this work that the volatile content of the planetesimals (or pebbles) that are destroyed while traversing the envelope is what remains well mixed in the envelope, thanks to ice sublimation. We assume as well that this volatile-component is entirely made of water. This is justified mainly by two reasons. First, as we mentioned before, water is thought to be the main volatile molecule present in planetesimals. Second, there is no accurate EOS available in the literature for an arbitrary mixture of volatiles that covers the large ranges of temperature and pressure present in planetary interiors. The only EOS suited for this purpose is that of water. 

Thus, we adopt an EOS for a mixture of H, He and H$_2$O which takes into account degeneracy due to free electrons. For the H-He component we implement the \citet{SCVH} EOS, and for the H$_2$O component we use an improved version of ANEOS \citep{aneos}. An important drawback in the standard version of ANEOS is the assumption of the substance to be monoatomic in the gas phase. We have corrected this for water \citep{Benitez16} to include the proper degrees of freedom, following the approach of \citet{Melosh07}. We have implemented this new version of ANEOS for our water component.
The EOS of the mixture of H-He with water is obtained, as in \citet{Baraffe08}, by means of the additive volume rule, which has been proven to yield adequate results for mixtures of H, He and H$_2$O \citep{Soubiran15}.
 
\subsection{Accretion rate of solids}\label{AccretionRates}
We adopt different models for the accretion rate of solids:
\begin{enumerate}
\item \textbf{Constant accretion rate.} This basic scheme allows us to analyse the effect of envelope enrichment in its simplest form. For nominal results, we adopt a $\Mzdot = 10^{-6}$ $\Mearth$/yr. Other values of accretion rates are tested in Sect. \ref{other_Mzdot}.
\bigskip

\item \textbf{Accretion rate \textit{a la Pollack}.} We implement our enrichment code in a Pollack-scheme \citep{P96} planetesimal accretion rate (see Sect. \ref{P96_section} for an explanation). The purpose of this is to show the effect of envelope enrichment in a more realistic formation scenario \citep[although still not very accurate, since the proper excitation of the planetesimals in the oligarchic growth regime is not included as it is in ][]{Fortier07,Fortier13}. The crucial parameters assumed for this scenario are summarised in Table \ref{paramP96}. The different choices of $\Sigma_0$ for the different sets of opacities are set to obtain similar formation timescales, as is explained later in Sect.\ref{P96_section}. For the capture radius we use the prescription of \citet{Inaba03} \citep[see as well,][]{Guilera11}. 

\begin{table}
\centering
\begin{tabular}{  l | c }
\hline\hline\
   Initial solid surface density ($\Sigma_0$) &  4 g/cm$^2$\\
   Planetesimal density ($\rho_{\text{p}}$) & 0.92 g/cm$^3$ \\
   Planetesimal radius ($r_{\text{p}}$) &100 km  \\
\hline  
\end{tabular}
\caption{Parameters used for results where the Pollack-scheme is implemented and nominal opacities (Mordasini + Freedman) are adopted. When we use opacities of Semenov + Ferguson the only parameter we change is the initial solid surface density, which is set at $\Sigma_0 = 10$ g/cm$^2$.}
\label{paramP96}
\end{table} 
\end{enumerate}

 \subsection{Boundary conditions}
For the definition of the radius of the protoplanet, we use the one proposed by \citet{Lissauer09}, which takes into account flow circulation from 2D hydrodynamical simulations: 
\begin{equation}\label{liss}
\Rp = \frac{G \Mp}{(c_s^2 + 4 G \Mp/ R_H)} ,
\end{equation}
$\Rp$ being the planet radius, $\Mp$ the planet mass, $c_s$ the sound speed of the gas  and $R_H$ the Hill radius. 

For nominal results, we adopt boundary conditions suited for the actual position of Jupiter. The values of these boundary conditions are given in Table \ref{tabBC}. The results are very insensitive to the boundary conditions (as long as the planet is beyond the ice-line and the accretion rate of solids is constant), this is why we do not show results for other boundary conditions. 

Regarding the choice of core density, we set the conservative number of 3.2 \rhocgs. This number should be larger once we consider that just rocky material sinks to the core, but we have tested that increasing the core density has a very small impact on the overall evolution.

\begin{table}
\centering
\begin{tabular}{ | l | c |}
\hline\
  $a $ & 5.2 AU \\
  $T_{\text{out}}$ & 150 K \\
  $P_{\text{out}}$  & 0.267 dyn/cm$^2$  \\
  $\rho_{\text{core}}$ & 3.2 g/cm$^3$\\
\hline  
\end{tabular}
\caption{Boundary conditions used in the nominal model. }
\label{tabBC}
\end{table}

\subsection{Choice of $\Mthresh$}\label{choice_mthresh}
The amount of material that is deposited in the envelope when a planetesimal is disrupted, is a complicated function that depends upon many planetesimal properties (mass, density, material strength, etc). Therefore, if one wants to compute self-consistently the mass deposition, an accurate knowledge of planetesimal properties is needed. Since this is still unknown, we encode this lack of information in the parameter $\Mthresh$, which represents the minimum mass of the core which can bind an envelope massive enough to destroy completely the incoming planetesimals. For nominal results, we take $\Mthresh$ = 2 \ME. We test the effect of considering, as well, $\Mthresh = 0.5, 1$, and 4 \ME $ $ in Sect. \ref{diffMthresh}. Just to have an order of magnitude in mind, a core mass of 1 \ME $ $ would have an envelope massive enough to disrupt completely, before reaching the core, icy planetesimals of $\sim 100$ m \citep{Podolak88}, a core of 2 $\Mearth$, icy planetesimals of $ \sim 1 - 10$ km, and a core of 4 \ME,  icy planetesimals of $\sim 10 - 100$ km \citep[Mordasini, priv. comm., based on calculations presented in][]{Mordasini06}\footnote{these values would be upper limits, planetesimals smaller that these would be fully disrupted in the envelope as well.}. A core of 0.5 \ME \, has an envelope mass equivalent to 2 Earth's atmospheres in our models, so pebbles would surely sublimate but not km-size planetesimals.

\section{Results for constant accretion rate of solids} \label{resultados}
\subsection{Differences in growth between standard H-He envelopes and enriched ones}\label{Mthresh2}
\begin{figure}
\begin{center}
\includegraphics[width=\columnwidth]{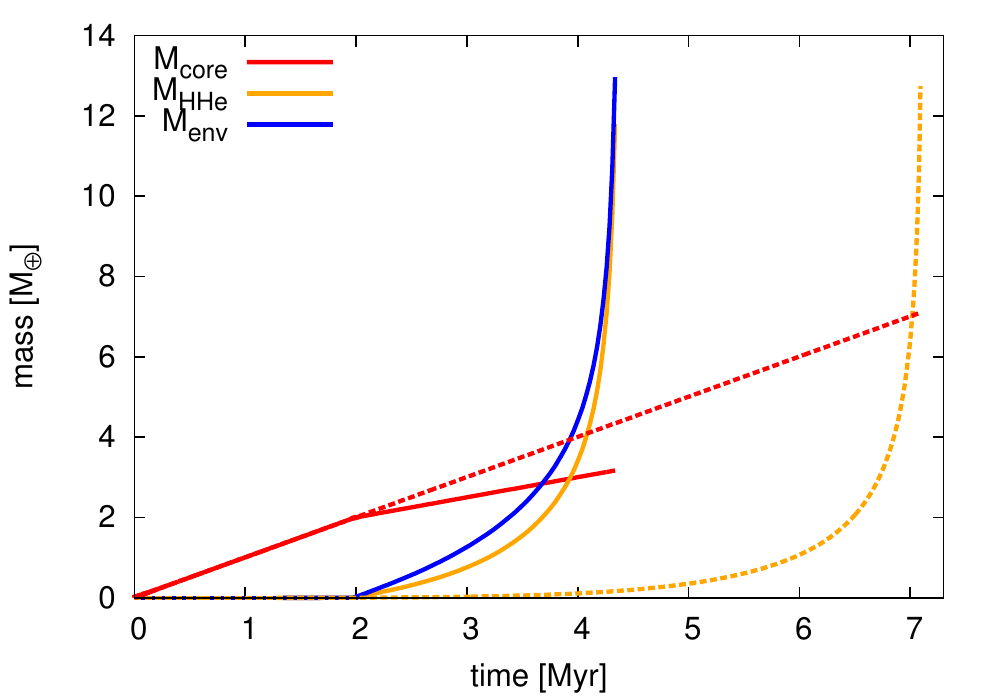}
   \caption{Solid lines: enriched case with $\Mthresh$ = 2 $\Mearth$ and $\beta$ = 0.5. Dashed lines: non-enriched case (H-He envelope).}
    \label{mass_Mzdot106}
\end{center}
\end{figure}

Figure \ref{mass_Mzdot106} shows the growth of a planet assuming $\Mzdot = 10^{-6} \Mearth/\text{yr}$ for a standard case where all the solids go to the core (envelope composed of H and He), and for a case where envelope enrichment is taken into account. Throughout this work, we will refer to the first case as ``non-enriched case" and the to the latter as ``enriched case". The first remarkable fact we observe is that the time to form a giant planet is reduced when envelope enrichment is taken into account. If we use the conservative definition of crossover time \citep[time when the core mass equals the envelope mass, $\tcross$,][]{P96} as the characteristic time to form a giant planet, we find $\tcross$ = 3.8 Myr for the enriched case against a $\tcross$ = 7.0 Myr for the non-enriched case. 

\begin{figure}
\begin{center}
\includegraphics[width=\columnwidth]{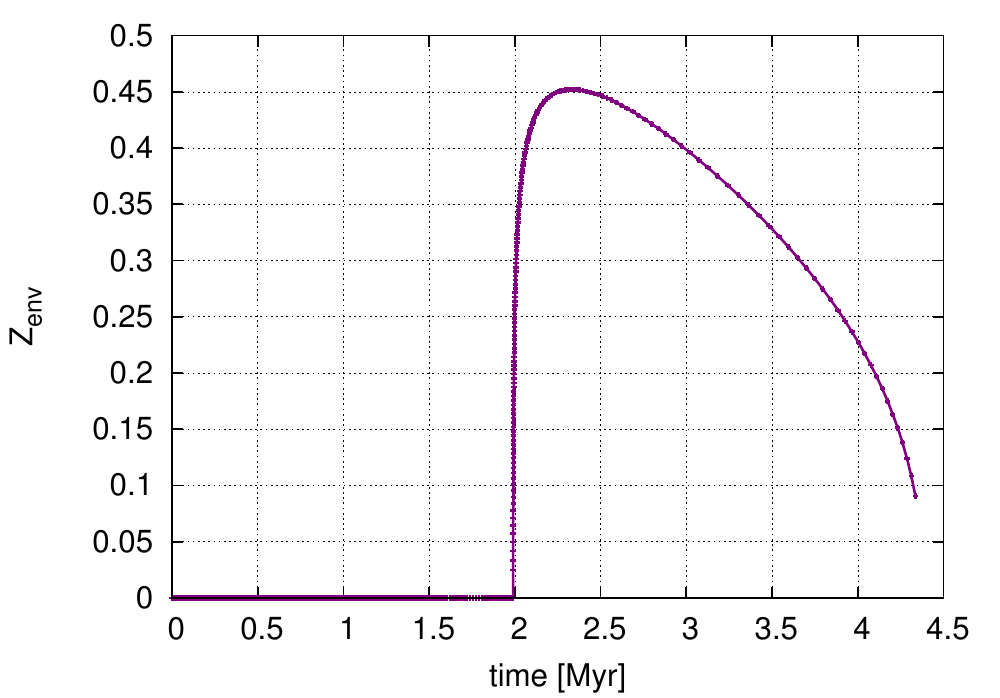}
   \caption{Evolution of the envelope metallicity for the enriched case of Fig. \ref{mass_Mzdot106}.}
    \label{Z_Mzdot106}
\end{center}
\end{figure}

\begin{figure}
\begin{center}
\includegraphics[width=\columnwidth]{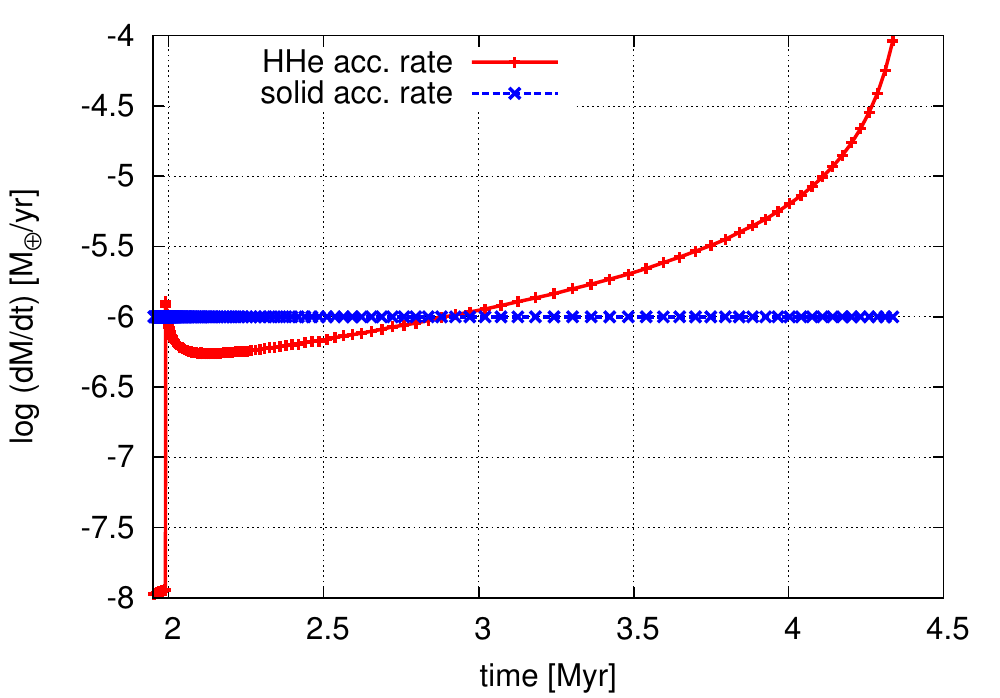}\\
\includegraphics[width=\columnwidth]{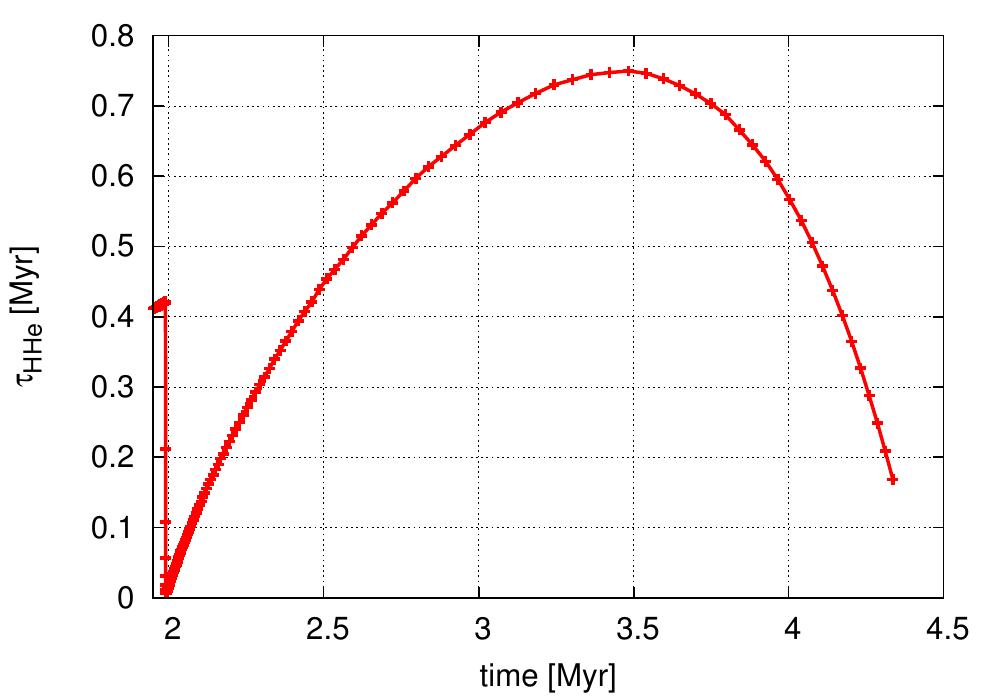}
   \caption{Top: accretion rate of H-He (red) vs. accretion rate of solids (blue) as a function of the core mass for the enriched case. Bottom: timescale to accrete H-He for the enriched case. Note that for $\Mcore \gtrsim$ 2.7 $\Mearth$, the accretion rate of H-He dominates the one of solids, and also the timescale to accrete H-He starts to decrease.}
    \label{AccRates}
\end{center}
\end{figure}

Let us analyse more deeply the enriched case, whose evolution of envelope metallicity is shown in Fig. \ref{Z_Mzdot106}. Once the core mass reaches a threshold value $\Mthresh$ = 2 \ME, half of the metals (water in our assumptions) remains uniformly mixed in the envelope and the other half is deposited in the core (see change of slope in $\Mcore$ in Fig. \ref{mass_Mzdot106}). Since the envelope is still quite thin at this $\Mcore$ (see Table \ref{depMthresh}), the envelope metallicity rises very rapidly at the beginning of the enrichment. In $3.5 \times 10^5$ yrs, $\Zenv$ reaches its maximum and then dilution slowly starts. There is an important fact to remark: rapid gas accretion is not triggered immediately as a consequence of envelope enrichment. This is well illustrated in Fig. \ref{AccRates}, where we observe that the accretion rate of H-He remains lower than that of metals until $t \sim 3 $ Myr . In the bottom panel of the same figure, we show the timescale to accrete H-He (defined as $\tauhhe =\MHHe/ \MdotHHe$) also as a function of time for the case of $\Mthresh = 2 $ \ME. We see that immediately after the onset of enrichment, the timescale to accrete H-He is very short, but this timescale starts to increase until it reaches a maximum at $t \approx 3.5$ Myr.

Analysing the timescale to accrete H-He allows us to infer when runaway of gas starts. As long as this timescale becomes larger, it means that gas accretion is slowing down. Conversely, once $\tauhhe$ starts to decrease, HHe-accretion accelerates. So we define the \textit{onset of the runaway of gas} as the time when $\tauhhe$ reaches its maximum,\footnote{Different definitions for the onset of runaway of gas have been given in the literature. Some define it  when $\tauhhe$ drops to a fraction of its maximum\citep{Piso14, Lee14}. Qualitatively, the crossover mass criterion \citep{P96} works to infer fast accretion of gas, but the truth is that after $\tauhhe$ starts to decrease, runaway of gas is inevitable (as long as there is gas left on the disk, of course).} and we denote this time as $\trun$.

\begin{figure}
\begin{center}
\includegraphics[width=\columnwidth]{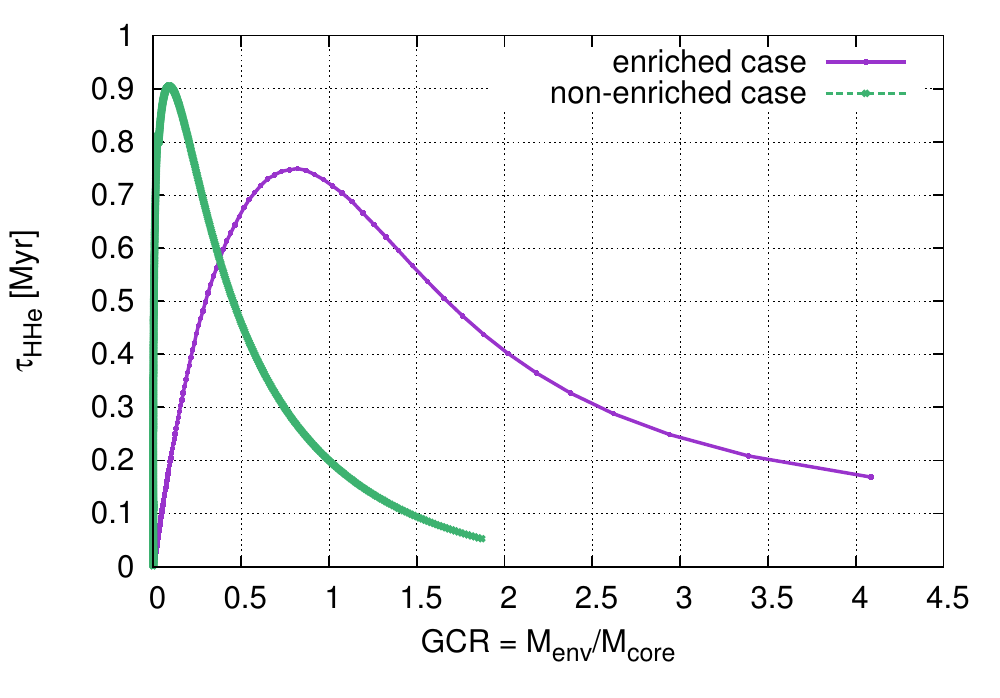}
   \caption{Timescale to accrete H-He as a function of the gas-to-core ratio, for the enriched case (purple) and the non-enriched one (green).}
    \label{GCR_tauHHe}
\end{center}
\end{figure}

The time elapsed between the onset of enrichment  ($t \approx 2$ Myr) and the onset of runaway of gas is 1.5 Myr. During that time, the envelope metallicity is always larger than 33\% (Fig. \ref{Z_Mzdot106}). This scenario implies that a very high gas-to-core ratio (GCR = $\Menv/\Mcore$) is achievable before runaway gas accretion is triggered. Fig. \ref{GCR_tauHHe} depicts this more clearly. It shows the timescale to accrete H-He as a function of the GCR. For the non-enriched case, the maximum of $\tauhhe$ occurs at GCR $\sim$ 0.1. For GCR larger than this, runaway of gas sets in. Therefore, protoplanets with GCR larger than 0.1 are expected to become gas giants. In other words, small and intermediate mass planets with high GCR are difficult to explain in the framework of the standard core accretion model, where envelopes are assumed to remain non-enriched \citep{Lee16}. Nevertheless, if envelope enrichment is taken into account, much larger CGR ($\sim 0.8$) can be achieved before the onset of runaway (Fig. \ref{GCR_tauHHe}). 

In this context we may wonder why are giant planets formed faster when envelope enrichment is taken into account.
When envelope enrichment sets in, the structure of the envelope changes radically: the mean molecular weight increases very rapidly due to the increase in $\Zenv$. Since the boundary conditions are the same, an increase of the mean molecular weight translates into an increase of the density profile. Therefore, the self-gravity of the envelope is stronger, and the planet becomes more prone to contract and accrete more gas.

\subsection{Dependence  on $\Mthresh$}\label{diffMthresh} 
In this section we test the effect of changing $\Mthresh$. As we stated in Sect. \ref{choice_mthresh}, this would correspond to consider different sizes (and/or composition, but mainly sizes) for the disrupted planetesimals/pebbles. Table \ref{depMthresh} summarises some aspects of the simulations for values of $\Mthresh$ = 0.5, 1, 2 and 4 \ME. The first row shows the mass of the envelope when envelope enrichment starts. This information is important when trying to link the value of $\Mthresh$ with the corresponding maximum size of planetesimals to be fully disrupted.

\begin{table}
\begin{tabular}{  l | c c c c  }
\hline\hline\
  $\Mthresh$ [\ME]&  0.5 & 1 & 2 & 4 \\
  \hline
  M$_{\text{env,thresh}}$ [\ME] & 2.6 $\times10^{-6}$ &1.35$\times10^{-4}$ & 4.9 $\times10^{-3}$ & 0.12 \\
 $r_{\text{P}}$ [km] & $\lesssim$ 0.001 & 0.1 & 1-10 & 10 -100\\
  $\Mcross$ [\ME] & 1.15 & 1.82 & 2.83 &  4.58 \\
  $\tcross$ [Myr] &  1.8 &  2.65 & 3.68 & 5.16  \\
  $\trun$ [Myr] &  2.67 &  2.9 & 3.48 & 4.83  \\
  ${\Delta t}_{\text{enriched}}$ [Myr] & 2.17 & 1.9 & 1.48 & 0.83 \\   
\hline  
\end{tabular}
\footnotetext{cm - m size particles}

\caption{Values of envelope mass at the onset of enrichment (M$_{\text{env,thresh}}$), approximate size of icy particles dissolved in the envelope for the corresponding $\Mthresh$, crossover mass  ($\Mcross$), crossover time  ($\tcross$), time at which runaway of gas begins ($\trun$) and time during which the envelope is enriched but still not in runaway of gas  (${\Delta t}_{\text{enriched}}$), for different values of $\Mthresh$. $\Mzdot$ = $10^{-6}$ $\Mearth$/yr. }

\label{depMthresh}
\end{table} 

\begin{figure}
\begin{center}
\includegraphics[width=1.05\columnwidth]{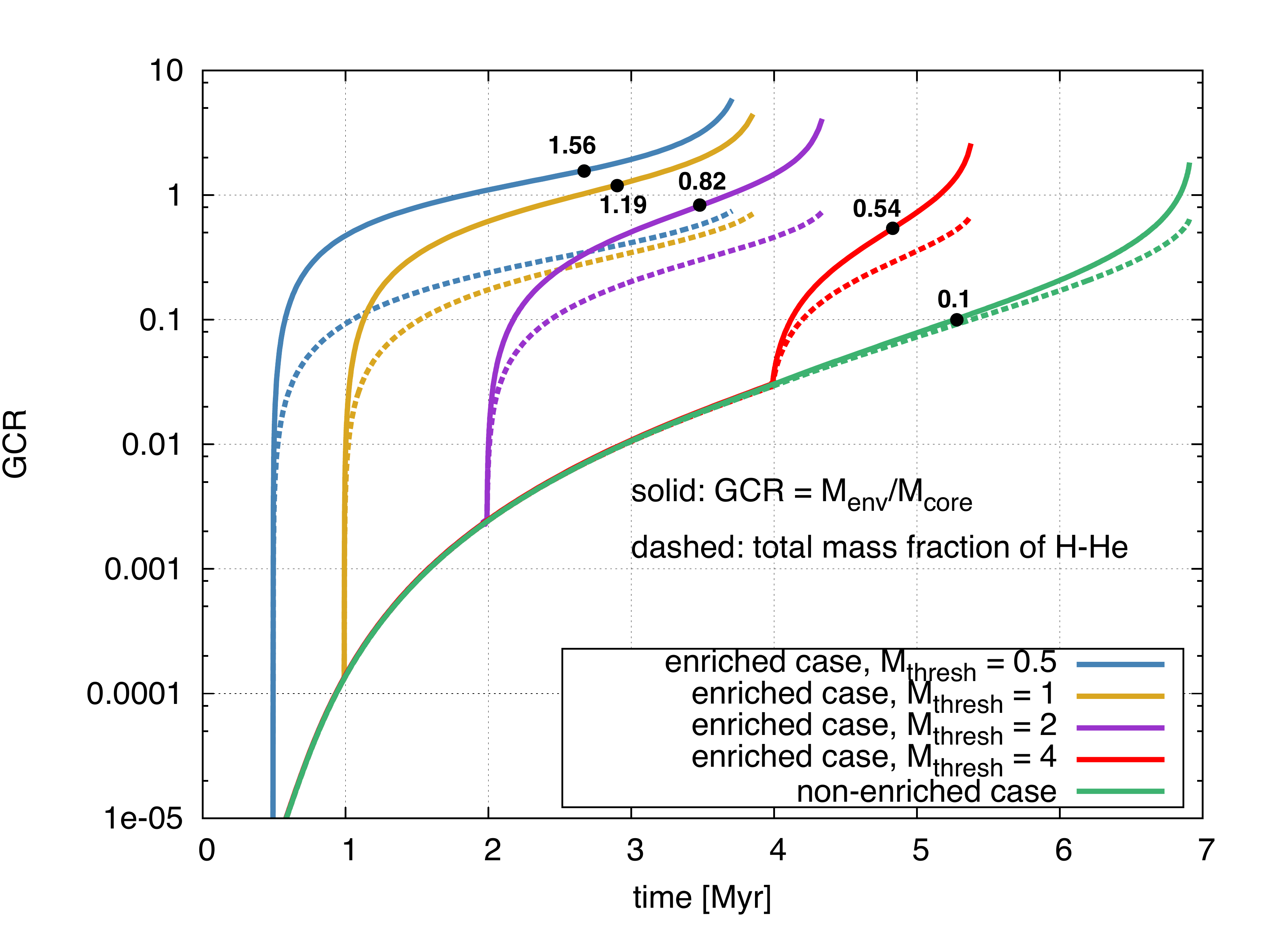}\\
\includegraphics[width=1.05\columnwidth]{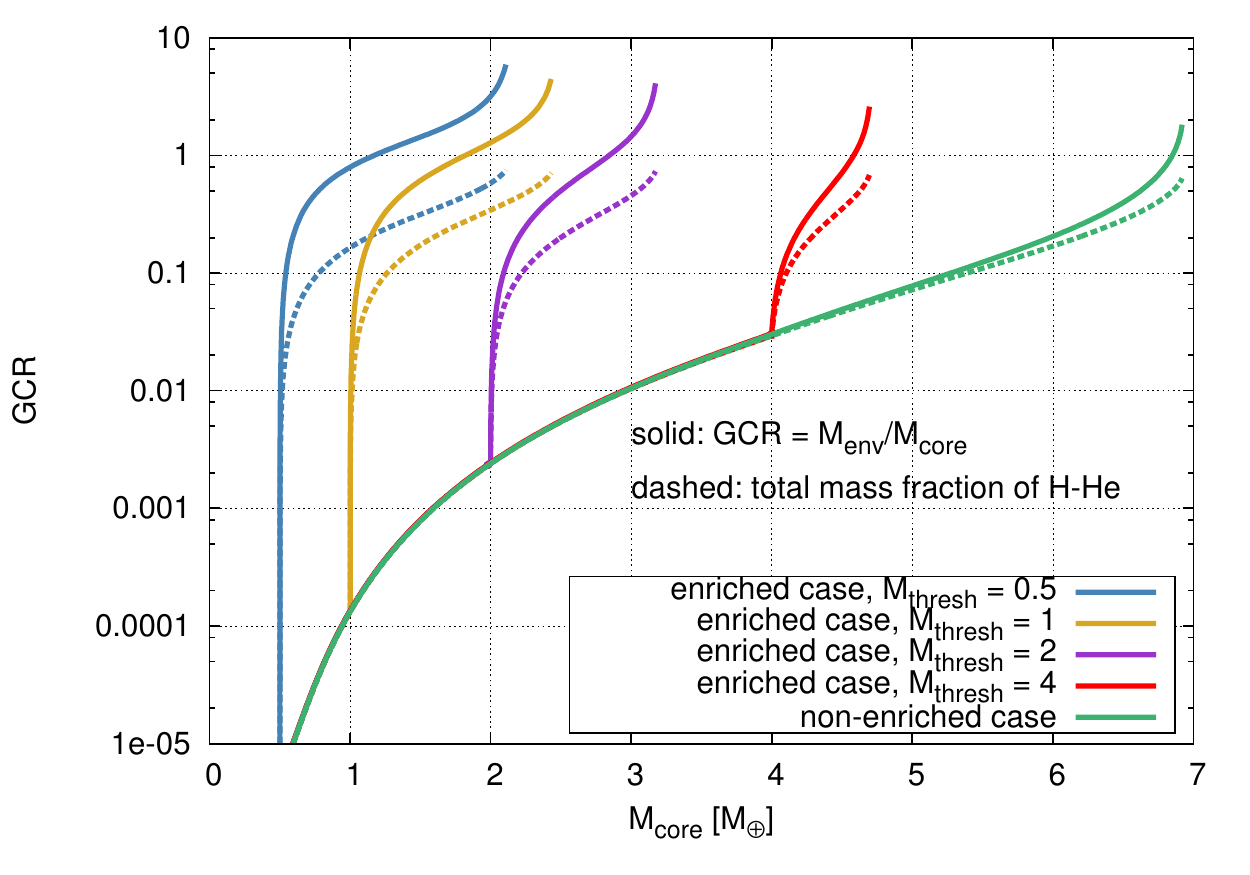}\\
   \caption{\textit{Top}: evolution of the gas-to-core ratio and total mass fraction of H-He of the planets for different $\Mthresh$ (in Earth masses, indicated with different colours at the inset of the figure). The black dots on each curve indicate the time when the runaway of gas starts, and the numbers above, the corresponding GCR at that time. \textit{Bottom:} same as above but as function of the mass of the core.}
    \label{manyMthresh}
\end{center}
\end{figure}

In Fig. \ref{manyMthresh} (top) we show the evolution of the gas-to-core ratio (GCR) and of the total mass fraction of H-He for the different $\Mthresh$. The black dots indicate the time of the onset of the runaway of gas (maximum in $\tauhhe$) and the numbers above, the corresponding value of GCR at this time.  

It is interesting to note that the smaller $\Mthresh$, the larger the gas-to-core ratio that can be achieved before runaway of gas.
Also interestingly, the period during which enrichment takes place but fast accretion of gas still does not occur, tends to be longer the lower $\Mthresh$ (last row of Table \ref{depMthresh}). This means that the smaller the planetesimals, the more likely it is that at the time the disk disappears, a small planet with high GCR remains. 

Another surprising fact is that the total mass fraction of H-He (which we denote f$_{\text{HHe}}$) \footnote{Note that the ``total metallicity" of the planet is $\Ztot$ = 1- f$_{\text{HHe}}$ = $(\Mcore + \Mzenv)/\Mp$} at the onset of runaway of gas is always $\sim$ 30\% for the enriched cases (see Table \ref{onset_runaway}), despite the value of $\Mthresh$. That value would be the maximum attainable for planets that do not become gas giants, smaller values of f$_{\text{HHe}}$ are also possible, as Fig. \ref{manyMthresh} depicts. This means that envelope enrichment seems to be a natural way to explain the formation of low mass, low density planets (mini-Neptunes) and also the recently called  ``super-puffs": light planets that have a bulk composition of H-He of presumably $\gtrsim$ 20 \% of H-He by mass \citep{Lee16, Lopez14}. We want to remark that the claim of envelope enrichment being a more natural scenario to explain the formation of objects with $\sim$ 20 \% of H-He is not just due to the fact that these values are reached before the onset of runaway of gas, but also, because these high amounts of H-He last \textit{longer} when envelope enrichment is included, as Fig.\ref{manyMthresh} shows.

\begin{table}
\begin{center}
\begin{tabular}{  l | c  c  c  c  c}
\hline\hline\
  $\Mthresh$ [\ME]&  0.5 & 1 & 2 & 4 & $\infty$ \\
  \hline
  GCR & 1.56  & 1.16  &  0.82  & 0.54  & 0.10\\
  f$_{\text{HHe}}$ &  0.35 &  0.31 & 0.30 & 0.28 & 0.09  \\
   $\Zenv$ &  0.44 &  0.43 & 0.33 & 0.18 & 0.00 \\
     $\Mcore$ [\ME] & 1.59 & 1.95 &  2.75 &  4.45  & 5.31 \\
  $\Mp$ [\ME] & 4.10 & 4.30 & 5.00 &  6.84  & 5.79 \\
\hline  
\end{tabular}
\caption{Values of gas-to-core ratio (GCR), total mass fraction of H-He (f$_{\text{HHe}}$, i.e 1-$\Ztot$), $\Zenv$, core mass ($\Mcore$) and total mass of the planet ($\Mp$) at the onset of the runaway of gas ($t=\trun$) for different $\Mthresh$ ($\Mthresh = \infty$ corresponds to the non-enriched case). $\Mzdot$ = $10^{-6}$ $\Mearth$/yr. }
\label{onset_runaway}
\end{center}
\end{table}
 
An interesting aspect in Table \ref{onset_runaway} is that the total mass of the planet at the onset of runaway of gas grows with increasing $\Mthresh$ but decreases for $\Mthresh = \infty$ (the non-enriched case). This is related to the fast accretion of H-He that is triggered once enrichment sets in.  If we analyse the core masses of the different $\Mthresh$ at the onset of runaway (Table \ref{onset_runaway}), we note that for $\Mthresh = \infty$ the core is larger than for $\Mthresh = 4 \, \Mearth$. However, because the mass fraction of H-He is much larger in the enriched cases that in the non-enriched one, the total planetary mass of the $\Mthresh = 4 \, \Mearth$ case is larger than the $\Mthresh = \infty$ one.

\begin{figure}
\begin{center}
\includegraphics[width=0.98\columnwidth]{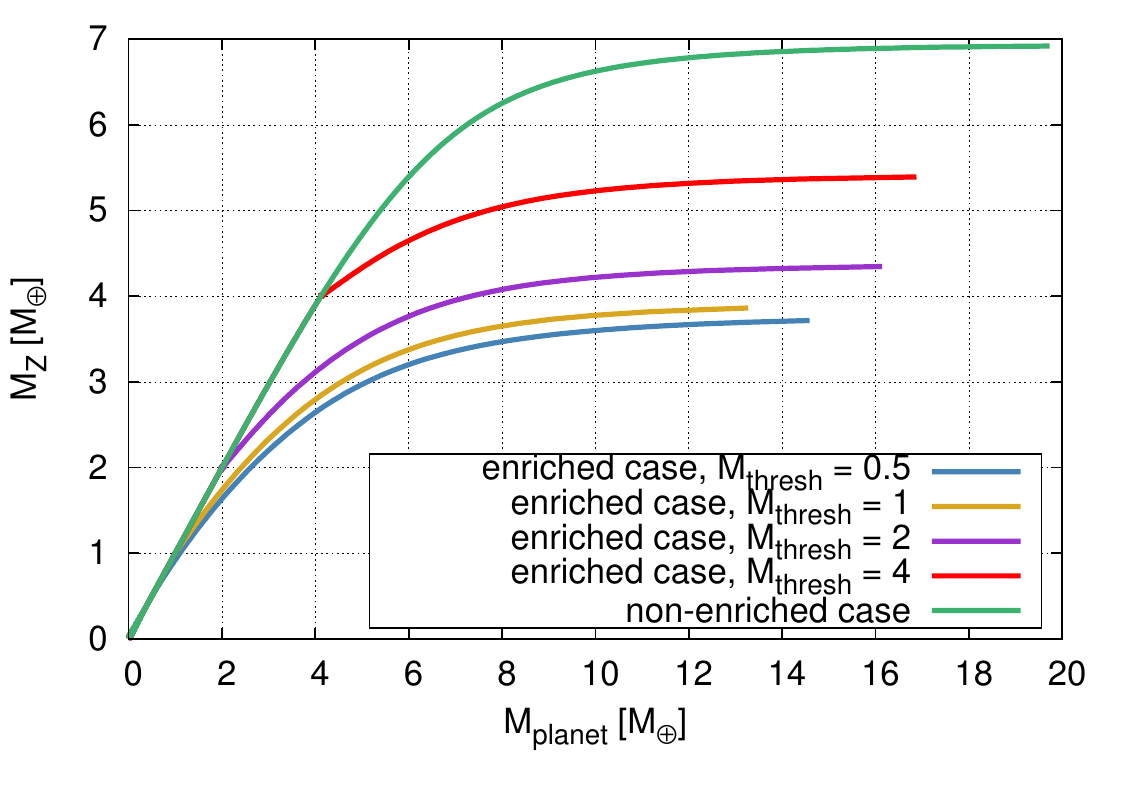}
   \caption{Total mass of solids ($M_Z$ = $\Mcore$ + $\Mzenv$) as a function of total planet mass ($M_{\text{planet}} = M_Z + \MHHe$) during formation. Note that simulations typically stop when $M_Z$ becomes asymptotically flat with planet mass. This flatness occurs because the accretion rate of H-He at this stage is much higher than the one of solids (usually by 2 orders of magnitude). We will use this fact to extrapolate total metallicities in Fig. \ref{Z_Mp} (just by fixing the final value of $M_Z$ in the definition of $\Ztot$).}
    \label{Mz_Mp}
\end{center}
\end{figure}

Finally, in Fig. \ref{Mz_Mp}, we show the total mass of solids of the planet as a function of the planetary mass acquired during growth. Note that with the low opacities of our nominal model (Mordasini + Freedman), even in the extreme case of non-enrichment, the maximum mass of solids that can be attained during growth is $M_Z \approx $ 7 $\Mearth$. Larger accretion rates can lead to larger core masses (or total mass of solids, in general). We show this in Sect. \ref{other_Mzdot}.
 
\subsection{Smoothing the transition in $\beta$}
One could think that since in reality there should be a distribution of planetesimal's sizes, then there is not one exact value of $\Mthresh$ for which the envelope starts to get enriched, but rather a range of $\Mthresh$. We have already tested the effect of adopting different $\Mthresh$ in the previous section, but considering a distribution of planetesimal sizes also means that $\beta$ would not change abruptly at a given $\Mthresh$, but it would transition for a range of $\Mthresh$.

\begin{figure}
\begin{center}
\includegraphics[width=0.9\columnwidth]{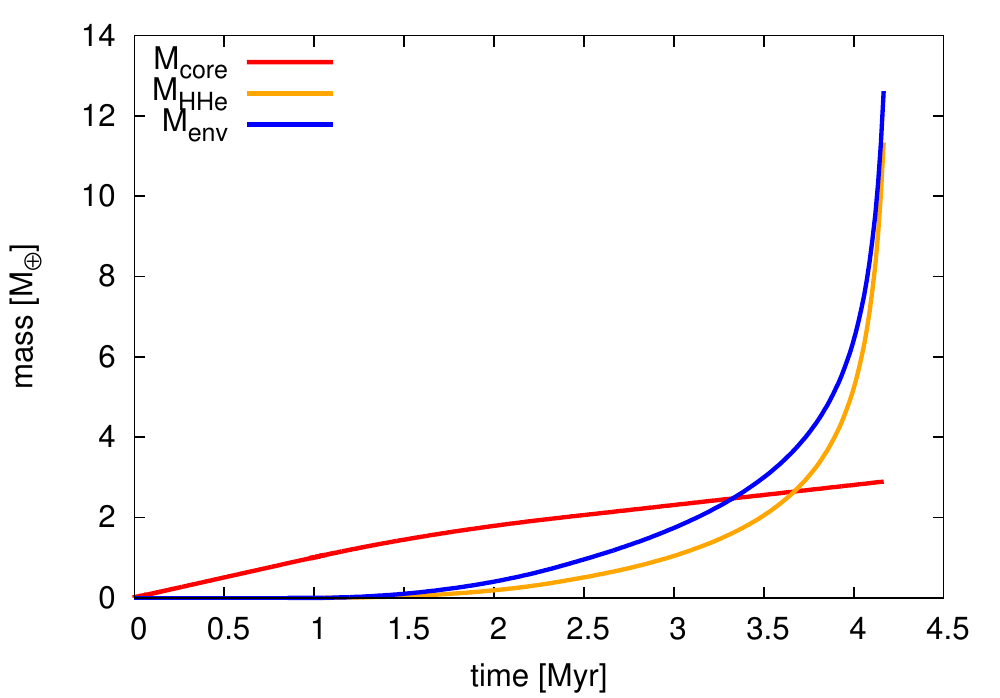} \\
\includegraphics[width=\columnwidth]{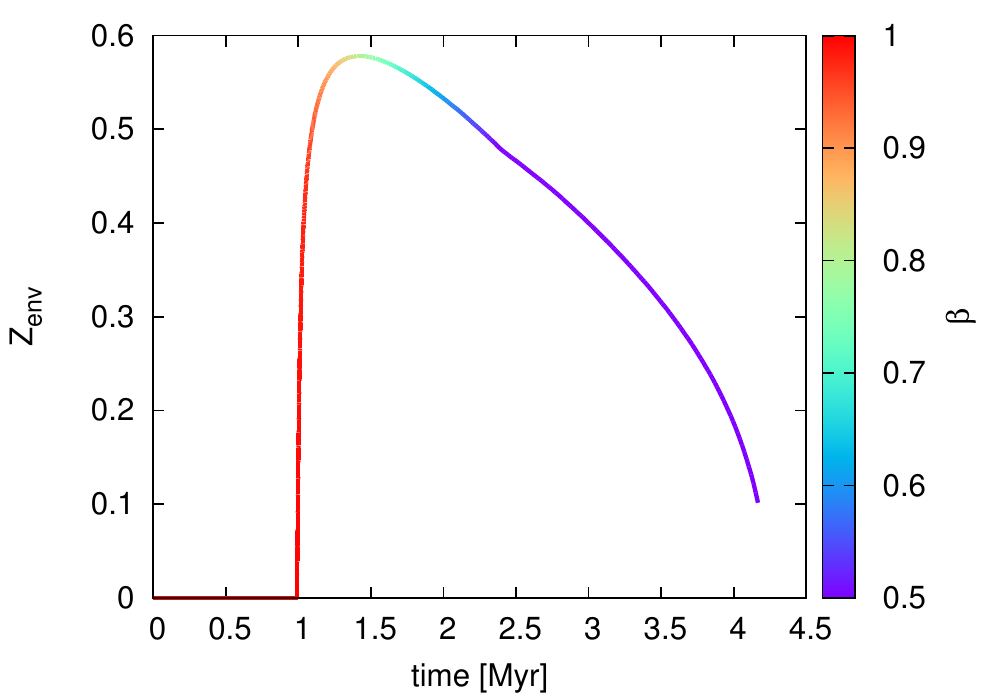}
   \caption{Evolution of masses and metallicity for a test case where we assume $\beta$ varying linearly between 1 and 0.5 for core masses ranging between 1 and 2 $\Mearth$. This smooth change of $\beta$ is shown in color-bar with the evolution of $\Zenv$.}
    \label{smoothbeta}
\end{center}
\end{figure}

We perform, therefore, the following test. We assume $\beta =1 $ for $\Mcore \le 1 \Mearth$,  $\beta =0.5 $ for $\Mcore \ge 2$ $\Mearth$, and a linear variation of $\beta$ in between. The growth of the planet is shown in Fig. \ref{smoothbeta}, as well as the evolution of envelope metallicity. Note that despite the smoother transition in $\beta$ than in the previous cases, the evolution is very similar to the case  $\Mthresh = 2$ $\Mearth$. The envelope metallicity still grows quite abruptly, because the envelope at $\Mcore$ = 1 $\Mearth$  is quite thin. Since the envelope at  $\Mcore$ = 1 $\Mearth$ is thinner than at  $\Mcore$ = 2 $\Mearth$, in this case of smoother transition of $\beta$, $\Zenv$ reaches a higher maximum than in the case $\Mthresh = 2$ $\Mearth$. But the overall formation time is practically the same as the case $\Mthresh = 2$ $\Mearth$. Therefore we conclude that assuming a sharp transition of $\beta$ at a given $\Mcore$ is not a relevant simplification.

\subsection{Other opacities}\label{BLFerg_Zenv}
As we mentioned in Sect. \ref{opac_paper2}, the timescale to form giant planets depends strongly on the choice of opacities. Despite the fact that in our nominal results we used the latest opacities published, it could be that the opacities do not follow exactly those low values. For instance, the recondensation of upstreaming gas into grains would increase the opacities, and this is not taken into account for the computation of \citet{Mordasini14} nor \citet{Freedman14}. 

Just to test how choosing larger opacities would affect our results, we run a simulation with the other extreme set of opacities that we introduced in Sect. \ref{opac_paper2}: dust opacities from \citet{Semenov03} and gas opacities from Ferguson (based on  \citet{AlexFerg94}, but including arbitrary metallicities). We consider here the enriched case with $\Mzdot = 10^{-6}$ $\Mearth$/yr and $\Mthresh =$ 2 $\Mearth$.

\begin{figure}
\begin{center}
\includegraphics[width=\columnwidth]{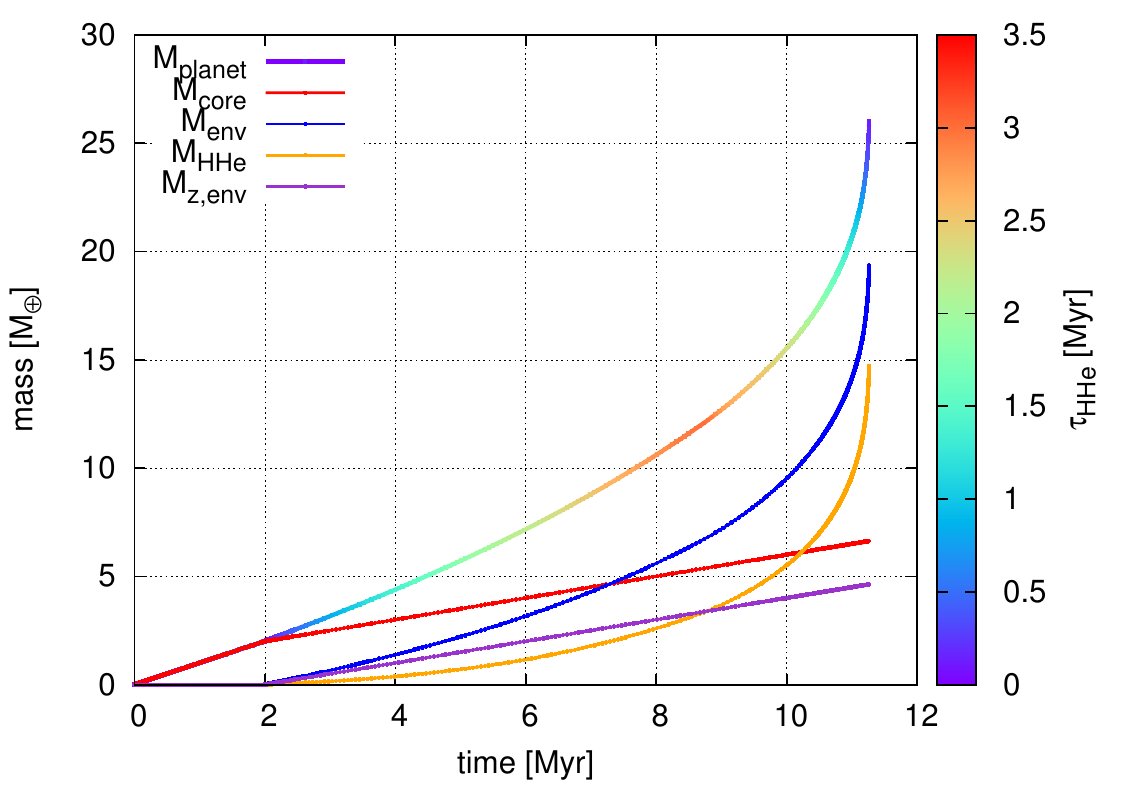} 
   \caption{Planet growth with Semenov + Ferguson opacities (see Sects.\ref{opac_paper2},\ref{BLFerg_Zenv}). $\Mthresh = 2$ $\Mearth$, $\beta = 0.5$ and $\Mzdot = 10^{-6}$ $\Mearth$/yr. } 
    \label{kappaBLFerg_masses}
\end{center}
\end{figure}

The results of the growth of the planet are shown in Fig. \ref{kappaBLFerg_masses}. The total mass of the planet is plotted with a color-bar that indicates the timescale to accrete hydrogen and helium ($\tauhhe$). We note that the maximum of this timescale in this case is reached after crossover mass. It occurs for a total mass of $\sim 12$ $\Mearth$, of which $\Mcore \sim$ 5.4 $\Mearth$ and $\Menv \sim 6.6$ $\Mearth$. At that time, half of the mass of the envelope is H-He and the other half, water. It means that when $\tauhhe$ reaches it maximum ($t \approx$ 9.5 Myr), we have a planet with a total mass and total metallicity ($ \Ztot \sim$ 72 $\%$) similar to Uranus.

The timescale shown here to form an ice giant seems quite long for a typical disk's lifetime, but this is just a consequence of the choice of the planetesimal accretion rate. For instance, if instead of using $\Mzdot = 10^{-6}$ $\Mearth$/yr we chose $\Mzdot = 3 \times 10^{-6}$ $\Mearth$/yr, then $\trun \approx 4$ Myr, and the overall evolution in terms of mass and metallicities is quantitatively similar to the results shown in Fig. \ref{kappaBLFerg_masses}. In the case of higher accretion rate of solids, at $t = \trun$, the total planetary mass is that of Neptune, with a total metallicity of $\Ztot \sim 70 \%$. 
Hence, for the high opacities used in this section, larger accretion rates lead to shorter formation timescales, yielding still, as a typical output, a Neptune/Uranus-like planet.

\subsection{Other accretion rates of solids}\label{other_Mzdot}
For completeness, we test in this section the effect of considering other values of accretion rates of solids. We run simulations with envelope enrichment  and the microphysics of our nominal model (low opacities) but using $\Mzdot = 10^{-5}$ $\Mearth$/yr and $5 \times 10^{-5}$ $\Mearth$/yr instead of  $\Mzdot \times 10^{-6}$ $\Mearth$/yr.
We summarise results in Tables \ref{diff_acc_rates_crossover} and \ref{diff_acc_rates_trun}.

Increasing the accretion rate of solids increases the core luminosity. Thus, the larger the accretion rate of solids, the hotter the envelope, which prevents gas accretion. Therefore, by increasing $\Mzdot$, we expect to obtain planets with larger planetary masses before the onset of runaway of gas.
Indeed, we note that with an accretion rate as high as $5 \times 10^{-5}$ $\Mearth$/yr (more typical for pebble accretion than for planetesimal accretion), when $\tauhhe$ reaches its maximum, the total mass of the planet is $\Mp\approx$ 11 $\Mearth$, with a core of $\Mcore = 5$ $\Mearth$ and total fraction of H-He of $\sim 30 \%$. This means that this could be other mechanism to form intermediate-mass planets. Actually, this is exactly the mechanism proposed by \citet{Lambrechts14} to form Uranus and Neptune via pebble accretion. The problem with this scenario is that the formation timescale is very short ($\sim$ 0.2 Myr). So it is hard to justify why a planet that reached the onset of runaway of gas in 0.2 Myr would not continue accreting gas to become a gas giant. We can only think that the embryo was formed late in the disk, but then we fall in the classical fine-time tuning problem: the embryo has to form when the disk is fading, but before the gas is totally gone. Therefore, this scenario to explain the formation of Neptunes seems unlikely. We discuss further the formation of ice-giants in the framework of pebble accretion in Sect. \ref{pebbles}.

\begin{table}
\begin{center}
\begin{tabular}{  l | c  c  c  }
\hline\hline\
  $\Mzdot$  [\ME/yr] &  $10^{-6}$ &  $10^{-5}$ &  $5 \times 10^{-5}$   \\
  \hline
$\Mcross$ [\ME]  & 2.83 & 3.7 &  4.45 \\
$\tcross$  [Myr]  &  3.68 &  0.54 & 0.14 \\
 $\trun$ [Myr]  & 3.48 & 0.56 & 0.16  \\
\hline  
\end{tabular}
\caption{ Crossover mass, crossover time, and time of the onset of the runaway of gas for different accretion rates of solids. $\Mthresh$ = 2 \ME.}
\label{diff_acc_rates_crossover}
\end{center}
\end{table} 

\begin{table}
\centering
\begin{tabular}{  l | c  c  c  }
\hline\hline\
  $\Mzdot$  [\ME/yr] & $10^{-6}$  & $10^{-5}$ & $5 \times 10^{-5}$ \\
  \hline
 $\Mp$ [\ME]  & 5.0 & 7.8  &  11.3  \\
 $\Mcore$ [\ME]  &  2.75 & 3.77 & 5.1 \\
$\MHHe$ [\ME]   & 1.5 & 2.24 & 3.26 \\
 $\tauhhe$ [yr]  & $ 7.5 \times 10^{5}$ & $1.6  \times 10^{5}$ & $5.2 \times 10^{4}$ \\
\hline  
\end{tabular}
\caption{ Total mass, core mass, mass of H-He and value of $\tauhhe$ at the onset of runaway of gas (t = $\trun$) for different accretion rates of solids. $\Mthresh$ = 2 \ME.}
\label{diff_acc_rates_trun}
\end{table} 

\subsection{On water condensation}
In all the simulations presented in this work we have included the effect of water condensation (as it should be). In \citet{Venturini15} we showed that water condensation decreases the adiabatic gradient, and this plays a relevant role in diminishing the critical core masses to very low values, especially for very high envelope metallicities ($\Zenv \gtrsim 0.6$). 

We have repeated some of the simulations presented before, but not including the effect of water condensation in the computation of the adiabatic gradient. For doing this, we run simulations using CEA  for the equation of state \citep{CEA} as in \citet{Venturini15}, because with this package it is possible to switch off the effect of water condensation in the computation of the adiabatic gradient. The results presented in this section were, therefore, all performed with CEA (those where water condensation is included, and those were it is not). 

In the nominal case where we use Mordasini + Freedman opacities, the difference between including or not water condensation does not really affect the evolution, because with these low opacities the outer layers of the envelope are radiative, so the structure in those layers is independent of the value of the adiabatic gradient.

For higher opacities and/or accretion rates of solids, the envelopes are more prone to be convective, so for those cases, water condensation plays an non-negligible role. For instance, considering the nominal case but increasing the dust opacities by a factor of 300, and taking an accretion rate of solids of $\Mzdot $=  3 $\times 10^{-6}$ $\Mearth$/yr, we find that the total formation time (when the planet reaches a mass of $\sim$ 40 $\Mearth$) is of 5 Myr for the case where water condensation is taken into account, and 6 Myr when it is not. Concerning the maximum $\Zenv$ attained, for the former case it is 0.75, and for the latter, 0.85. This is consistent with what we found in \citet{Venturini15}: water condensation makes envelopes thicker for the same core mass, which explains the fact of reaching a smaller envelope metallicity and accreting gas faster than in the case where water condensation is neglected.

\section{Results with an accretion rate of metals as in P96}\label{P96_section}
In this section we study the effect of envelope enrichment in the framework of a \citet{P96} accretion scheme. The main difference with fixing a given accretion rate of planetesimals is that now the accretion rate of planetesimals depends on the total mass of the protoplanet and that the availability of planetesimals to be accreted depends on the initial amount at the neighbourhood of the embryo (the initial surface density of solids). 

We implement first an accretion rate of solids \textit{a la Pollack} with initial surface solids of $\Sigma_0$ = 4 g/cm$^2$  and nominal opacities (Mordasini + Freedman). It is important to note that the value of $\Sigma_0$ is arbitrary and has been chosen so as to obtain a formation timescale for the non-enriched case of 8 Myr, as in \citet{P96}. However, in our calculations this value has no influence on the opacities even though it is likely that dust-to-gas ratio and total disk mass would change opacities. 
Recent results by \citet{Mordasini14} and \citet{Ormel14} tend to show that a least regarding the dust, an increase in the accretion rate of solids (or $\Sigma_0$ in the context of the P96 accretion scheme) does not increase the dust opacity, because the larger the flux of dust received by the protoplanet, the faster the coagulation and settling within the envelope.

\begin{figure}
\begin{center}
\includegraphics[width=\columnwidth]{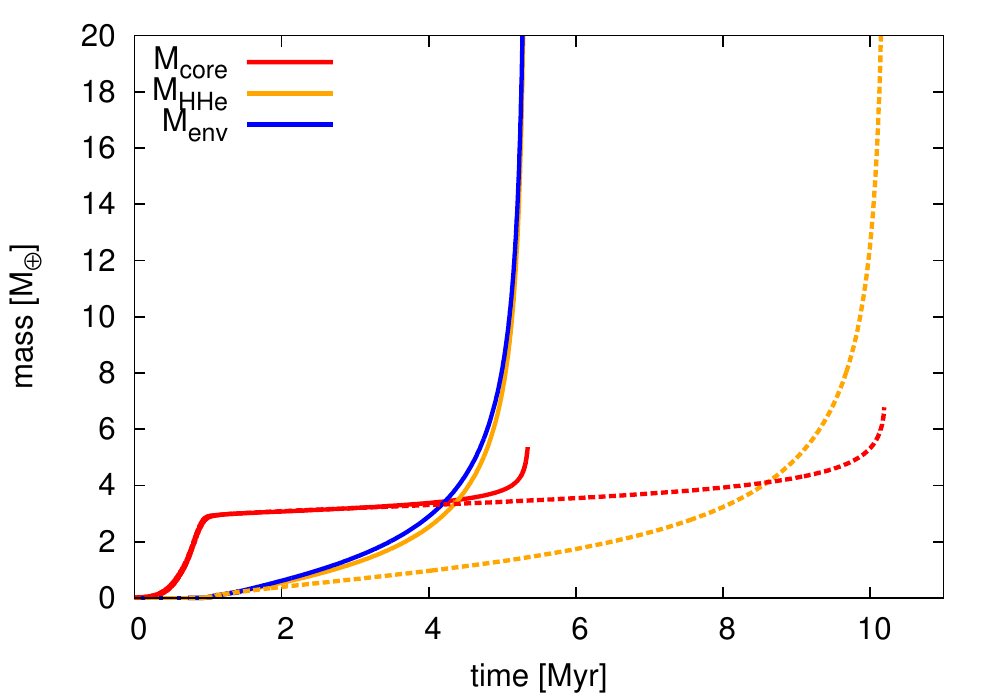} \\
\includegraphics[width=\columnwidth]{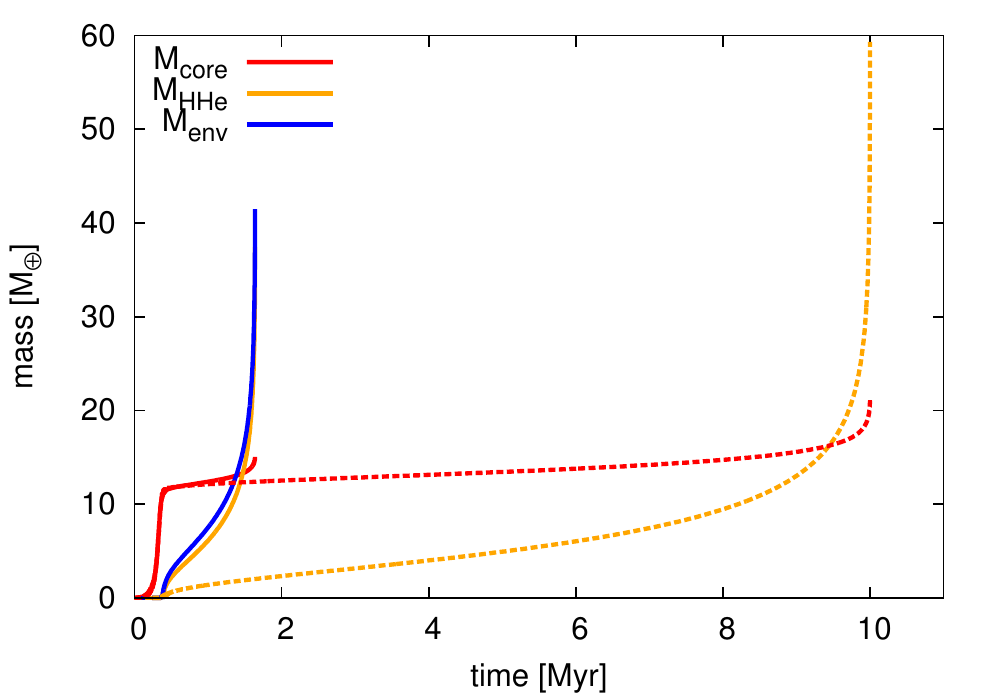}
   \caption{Growth of the planet using a P96 accretion scheme, with the assumption of accretion of icy planetesimals of 100 km size. For the enriched cases, $\Mthresh$ was chosen, for self-consistency, to correspond to the breakup of icy planetesimals of 100 km size. The solid lines correspond to the enriched case, and the dashed lines to the non-enriched case (classical case where the envelope is composed of H-He). \textit{Upper panel}: nominal opacities (Mordasini + Freedman). The initial surface density of solids is $\Sigma = 4$ g/cm$^2$. \textit{Lower panel}: high opacities (Semenov + Ferguson). The initial surface density of solids is $\Sigma = 10$ g/cm$^2$. Note that in the high-opacity case, when envelope enrichment is taken into account, the overall formation time is reduced by a factor of $\sim$ 6. The difference on the initial surface density values between the two figures was chosen to get the same overall time formation for the non-enriched cases.}
    \label{P96mass}
\end{center}
\end{figure}

The results for the mentioned  opacities and surface density of solids are illustrated in the upper panel of Fig. \ref{P96mass}. The enriched case assumes $\Mthresh$ = 3 $\Mearth$ and $\beta$ changing at this core mass from 0 to 0.5. The choice of $\Mthresh$ is to ensure that the envelope mass is the appropriate to dissolve icy planetesimals of 100 km radius (which is the choice of planetesimal size for the P96 scheme, see Tab.\ref{paramP96}). In this case, when phase 1 ends, the core mass is 2.9 \ME, which means that $\Mthresh$ is reached during phase 2 of P96. We note that the decrease in formation time, compared to the non-enriched case, is of a factor $\sim$ 2.

We test as well the effect of using the combination of high opacities (Semenov + Ferguson) in this P96 accretion scheme (Fig. \ref{P96mass}, bottom). To get the same overall formation time as before, for the non-enriched case, we choose an initial surface density of solids of $\Sigma = 10$ g/cm$^2$ (as in the baseline model of P96). Because of the high opacities and the high accretion rate of solids achieved in phase 1, the envelope required to destroy icy planetesimals of 100 km size corresponds now to $\Mthresh = 11 $ $\Mearth$. We note that in this case, the formation timescale is reduced by a factor $\sim$ 6 when envelope enrichment is included. The difference with the low-opacity case is that now $\Mthresh$ is achieved just before phase 1 ends. This means that when envelope enrichment takes place during phase 1, there are feedback mechanisms acting that favour more rapid gas accretion. In order to understand these mechanisms, we proceed to analyse the behaviour of the solid accretion rate for the high-opacity scenario.

\begin{figure}
\begin{center}
\includegraphics[width=0.96\columnwidth]{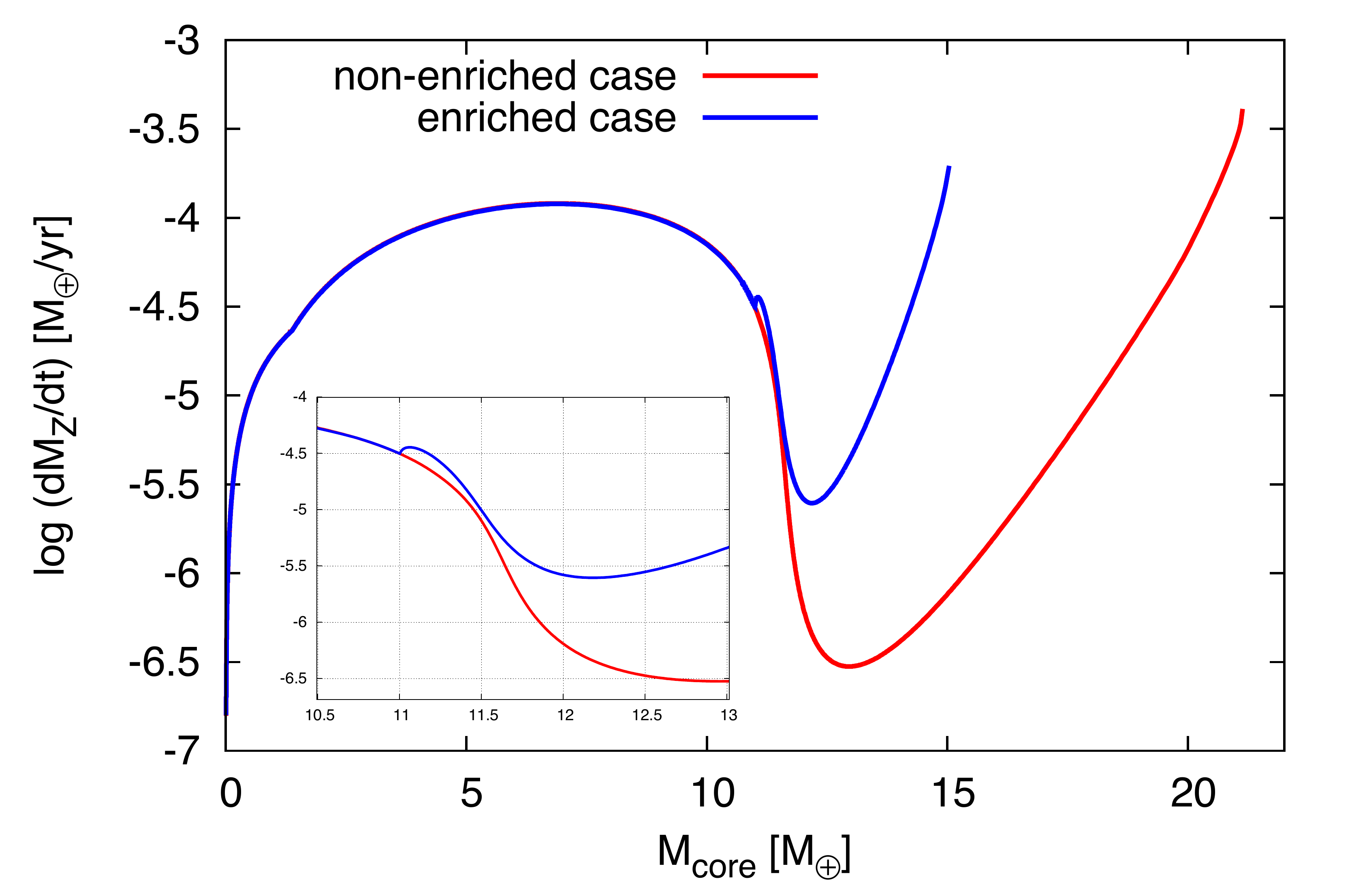} \\
\includegraphics[width=0.96\columnwidth]{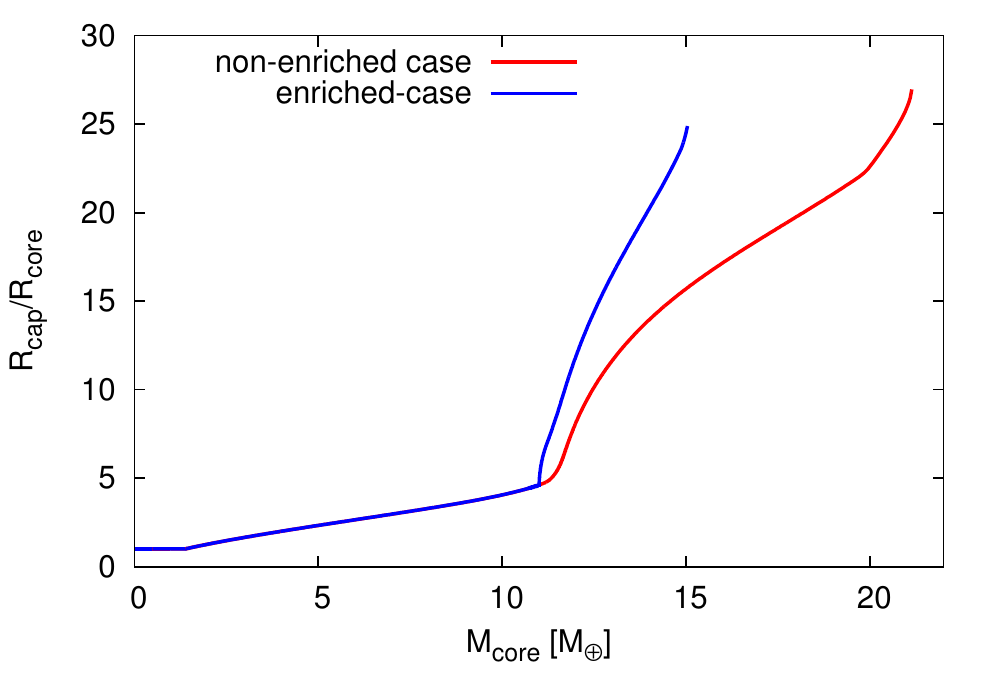}
   \caption{\textit{Upper panel}: $\Mzdot$ as a function of core mass for the enriched and non-enriched cases of the high-opacities simulations (Fig. \ref{P96mass}, bottom). Note that the planetesimal accretion rate of the enriched case increases when the core reaches the threshold value of 11 \ME. \textit{Lower panel}: ratio of capture radius to core radius as a function of core mass (same simulation as in the upper panel).}
    \label{P96Accrates}
\end{center}
\end{figure}

The upper panel of Fig. \ref{P96Accrates} shows the change in the accretion rate of solids as a function of time for the enriched and non-enriched cases of the high-opacitiy simulation (Fig. \ref{P96mass}, bottom). A bump in the accretion rate of solids occurs once the envelope begins to be enriched (at $\Mcore = 11$ \ME). If we observe the evolution of the capture radius (Fig. \ref{P96Accrates}, bottom) we see that also at this moment, it grows considerably. While the outer boundary conditions are the same as in the non-enriched case, the mean molecular weight of the envelope increases as accretion proceeds. Therefore, the density increases, so the planetesimals are more efficiently slowed down when crossing the atmosphere. This increases the capture radius which translates into a larger accretion rate.

Note that, in principle, the same effect occurs if enrichment sets in during phase 2. However, in that case, the accretion rate of solids is much smaller, and the availability of the protoplanet to increase the accretion rate of planetesimals is limited to the amount of planetesimals that can enter in the feeding zone in each timestep. Hence, it makes sense that the increase of the accretion rate of solids is smaller than when enrichment starts in phase 1.

\section{Implications on the formation of different types of planets}\label{implications}
\subsection{Formation of gas giants}
We have shown that by including the effect of envelope enrichment during the growth of a planet, the timescale to form a gas giant is shorter than in the standard case of no envelope enrichment. In the classical context of planetesimal accretion, the formation of gas giants is challenging when the oligarchic growth regime is taken into account \citep{Fortier07, Fortier13} and even more challenging if planetesimal fragmentation is included \citep{Guilera14}. 

\citet{Fortier13} showed (for the classical scenario of H-He envelopes) that the formation of giant planets required small planetesimals ($\sim$ 100 m). They found that km-size planetesimals (and larger) were not sufficiently damped by gas drag, and therefore, due to their large relative velocities, were not efficiently accreted by the protoplanet. We have shown that envelope enrichment reduces formation timescales. Therefore, including this effect in simulations that consider oligarchic growth might help the formation of gas giants from km-size planetesimals. The reason for this is not related to envelope enrichment playing any role in the damping of planetesimals eccentricity, but to the fact that it accelerates the growth of the planetary envelope, producing a positive feedback in the increase of the capture radius. We showed in Sect. \ref{P96_section} that envelope enrichment can reduce by a factor of $\sim$ 5 the timescale to form a giant planet. The same effect should reduce the formation time of giants when oligarchic growth is taken into account. This scenario should be tested in the future.

In the framework of pebble accretion, the problem with gas giants is the opposite than with planetesimals: giant planets are extremely easy to form. In this context, envelope enrichment would make things even worse. But it is important to stress that the current status of pebble accretion (concerning giant planet formation) is that the embryos must be arbitrarily allowed to start to grow after the disk is significantly evolved \citep{Bitsch15}. Otherwise, most of the embryos growing beyond the iceline end up as gas giants. The feasibility of pebble accretion to reproduce (statistically) the giant planets observed needs still to be proven, and envelope enrichment should be included for those models to be physically acceptable. 

\subsection{Formation of mini-Neptunes and Neptunes}
Another important result we obtained, is that if the disk disappears before the onset of runaway of gas, the type of planets we can form with envelope enrichment is quite different than in the non-enriched counterpart. With envelope enrichment we expect failed giants to have large gas-to-core ratios ($\sim$1), high mass fraction of H-He  (up to $\sim$ 30\%) and relatively high envelope metallicities ($\gtrsim$ 40 \%). This result is very interesting, because one ongoing problem with the classical picture of the core accretion model (without envelope enrichment) is the formation of low ($\sim$ 1 -10 $\Mearth$) and intermediate mass ($\sim$ 10 - 20 $\Mearth$) objects with important contents of H-He and high total metallicity. This has been, for instance, a recurrent problem in the formation of Uranus and Neptune in the solar system, because even if the disk disappears at the moment when the mass of the planet is in the appropriate range (15 -20 $\Mearth$), a core-dominated planet with $\sim 20 \%$ in mass of H-He is difficult to obtain \citep{HelledBod14}. The same regarding mini-Neptunes: the maximum gas-to-core ratios expected from simulations for H-He envelopes is typically $\lesssim$ 10\% \citep{Lee16}.

Our nominal model is able to reproduce mini-Neptunes, but not Neptunes. This is because of the low opacities. Gas opacities from \citet{Freedman14} include the effect of a changing metallicity (and therefore, increase with increasing $\Zenv$), but the dust opacities of \citet{Mordasini14} are still very low. Therefore, with these opacities, crossover masses are low (even without envelope enrichment, see Table \ref{opacHHe}). 
It is not possible with these low opacities to obtain the static structure of a $\sim$ 15 $\Mearth$ planet with a core of $\sim$ 10 (no matter which value of envelope metallicity). 
Nevertheless, a Neptune-type planet can be formed with envelope enrichment if larger opacities are invoked, as we showed in Sect. \ref{BLFerg_Zenv}. This shows the importance of keep improving opacity models. The formation of clouds could be one mechanism for larger opacities to exist in the envelope of protoplanets, and we have shown that water clouds can be present in the atmospheres of protoplanets formed beyond the iceline \citep{Venturini15}. So effort on the direction of including this physics on the computation of opacities should continue. 

\subsection{Formation of ice giants via pebble accretion}\label{pebbles}
In Sect. \ref{other_Mzdot} we showed that when considering large accretion rates, as the ones invoked in the context of pebble accretion, the formation of ice giants was difficult because formation times were too short. In principle, the situation could be improved considering larger opacities (as we showed in Sect. \ref{BLFerg_Zenv} for planetesimal-like accretion rates). However, we note that when we run the non-enriched case with the high opacities of Semenov + Ferguson and a solid accretion rate of  $5 \times 10^{-5}$ $\Mearth$/yr, runaway of gas is triggered at $\trun \approx 0.45$ Myr. This tells us that if we considered envelope enrichment with these high opacities, formation timescales of giant planets would still be too short (of the order of $\sim$ 0.1 Myr). 

\citet{Lambrechts14} claim that they can explain the formation of ice giants in the context of pebble accretion. Their argument goes as follows: pebble accretion rates are so high that they provide a means for increasing the critical core mass to very large values \citep{Stevenson82, PT99}, so a Neptune mass planet is in this context still sub-critical, and therefore will not accrete large amounts of gas. 

Moreover, they find the existence of an isolation mass for pebbles, which has values of $M_{\text{iso}} \gtrsim 20$ $\Mearth$ for distances from the star $a \gtrsim$ 5 AU. The isolation mass is smaller the closer the planet is to the star, so they claim that Jupiter and Saturn reached isolation mass, which caused the onset of runaway of gas \citep[halting solid accretion promotes gas accretion,][]{Ikoma00}; whereas Uranus and Neptune did not.

Indeed, \citet{Lambrechts14} show that they can get the correct structure of the four giant planets of the solar system (in terms of total mass and total metallicity). However, they do not show any time evolution. Despite of the increase of critical core mass, if this mass is reached in 0.1 Myr, fast gas accretion from this stage on will be inevitable.
Perhaps a way to avoid runaway of gas is to consider that planets accrete both pebbles and planetesimals: when pebble isolation mass is reached, the planet could remain accreting planetesimals that provide the necessary luminosity to delay runaway of gas \citep[just as in phase 2 of][]{P96}. This will be the subject of a future work.

\section{Predictions and comparison with observations}\label{predictions}
\subsection{Solar system}
We considered in this work enrichment by icy planetesimals. Therefore our predictions are relevant for planets formed beyond the iceline. Planetesimal disruption inside the iceline could take place as well, but given that the composition would be more silicate-iron rich, the thermal ablation of planetesimals would require either very small planetesimals' sizes, or thick envelopes. The last option would mean that enrichment would take place probably when the planet has already entered in the phase of runaway of gas. In addition, even if thermal ablation and mechanical disruption occur, the fate of a mixture of silicates with hydrogen is just expected to be miscible for $T \gtrsim \text{10,000 K}$  \citep{Wilson12}. Water has been proven theoretically to remain homogeneously mixed throughout the envelope of giant planets \citep{Soubiran15}, but for refractories, the fate of the fragments of planetesimals composed of those, is less certain, given the immiscibility just mentioned. If they tend to sink to the core, then the effect of envelope enrichment would not operate during formation, meaning that critical core masses would have the usual values of $\sim$ 10 - 20 $\Mearth$. This would imply that forming giant planets from rocky planetesimals (i.e, inside the iceline) would be difficult.

Regarding formation beyond the iceline, our results show that we can expect both gas giants and neptune-like planets to form via envelope enrichment, as explained above.

\begin{figure}
\begin{center}
\includegraphics[width=\columnwidth]{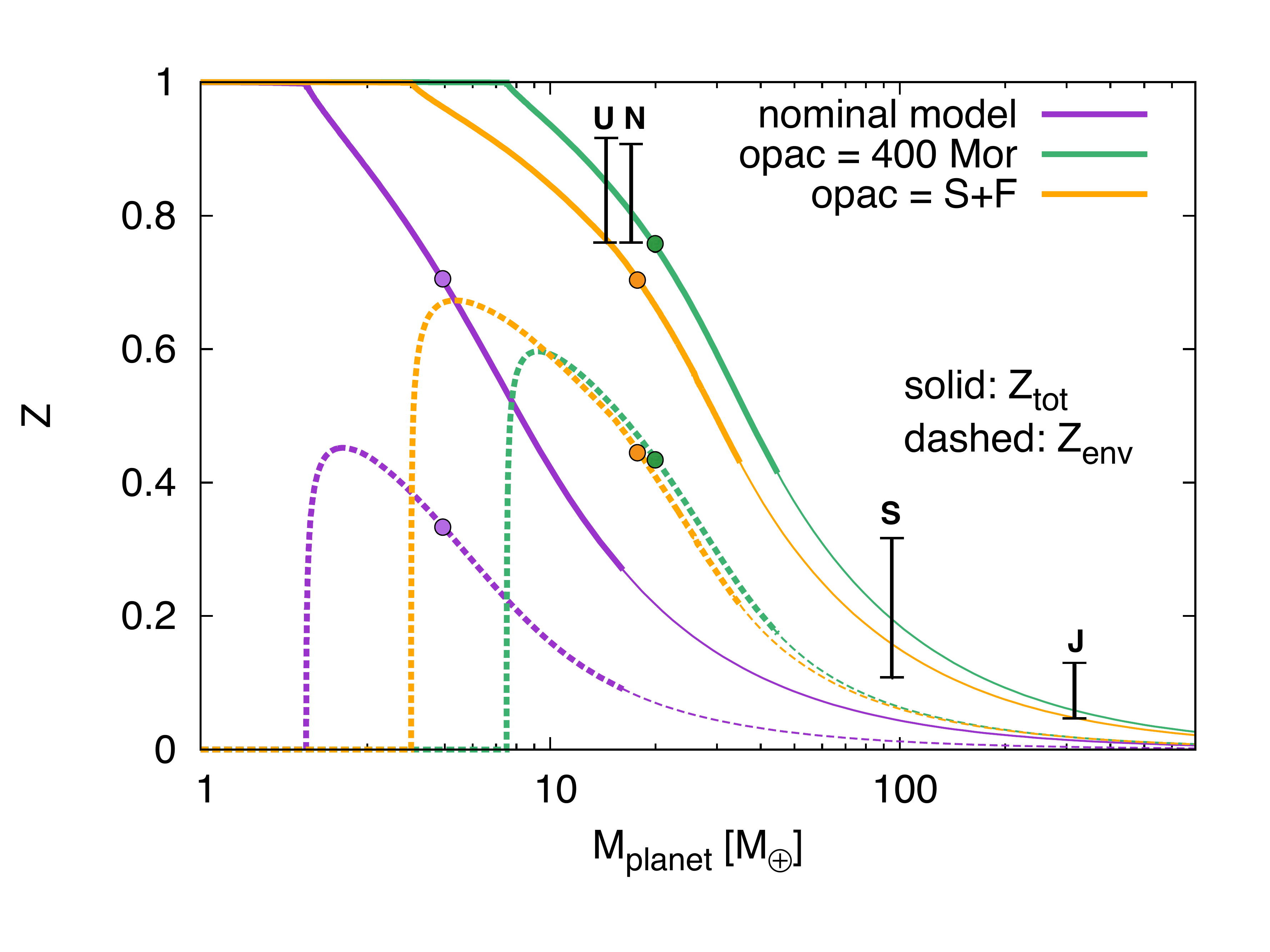} 
   \caption{Total metallicity (solid lines) and envelope metallicity (dashed lines) as a function of the planet mass for different models of planet formation with envelope enrichment. The different colours indicate different choices of opacities and accretion rates of solids. The violet corresponds to our nominal model (opacity of Mordasini + Freedman, and $\Mzdot = 10^{-6}$ $\Mearth$/yr). The orange uses opacities of Semenov + Ferguson, and  $\Mzdot = 3 \times 10^{-6}$ $\Mearth$/yr. The green curves were computed using the nominal opacities, but the dust opacities of Mordasini are enhanced by a factor of 400. The accretion rate of solids is $\Mzdot = 4 \times 10^{-6}$ $\Mearth$/yr. $\Mthresh$ was chosen such as the corresponding envelope is massive enough to destroy icy planetesimals of 1-10 km. The thick lines show the output of formation, and they all end at times of $\sim$ 4 - 5 Myr (the choice of the accretion rates was to get similar formation times). The thin lines are an extrapolation, and are shown just to see what we should expect in terms of metallicities up to jovian masses. The total metallicities predicted for the giant planets of the solar system are overplotted for reference (data taken from \citet{Helled11} for Uranus and Neptune, and from \citet{BaraffePPVI} for Jupiter and Saturn). The circular points indicate the onset of the runaway of gas. This means that masses at the right of these points, especially after the simulations stop, are unlikely (the expected elapsed time between the end of the simulations and the one required for $\Mp \approx$ 100 $\Mearth$ is $\sim 10^5$ yr ). This figure suggest that for the solar system, a combination of high opacities is preferred with respect to the low opacities adopted in the nominal model.}
    \label{Z_Mp}
\end{center}
\end{figure}

We have shown that by including envelope enrichment by icy planetesimals during formation, we expect the planets formed to present a decrease of envelope (and total) metallicity with increasing total mass. \footnote{The decrease of envelope metallicity with planetary mass was pointed out empirically by \citet{Kreidberg14}. The decrease of total metallicity with planetary mass is a natural outcome of core accretion \citep[e.g.,][]{Alibert05, Mord09}}.  
To illustrate this better, we show, in Fig. \ref{Z_Mp}, the output of planet formation for different choices of opacities and accretion rates of planetesimals. In all cases, the envelope starts to get enriched when it reaches a mass of $\Menv = 5 \times 10^{-3} \Mearth$, which corresponds to the fully disruption of icy planetesimals of approximately, 1-10 km-size (Sect. \ref{choice_mthresh}). We have overplotted the total metallicities of the giant planets of the solar system for reference. This figure suggest that rather high opacities are preferable to explain the giants of the solar system. Of course, the behaviour of the curves depends somehow on the choice of parameters. Still, when we analysed, for instance, the nominal model (Mordasini + Freedman opacities) for different $\Mthresh$, we saw that even the extreme case of non-enrichment could not lead to total high-Z content of more than $\sim$ 7 $\Mearth$ (Sect. \ref{depMthresh}). So it is clear that with those low opacities we cannot explain Jupiter, which has at least 15 $\Mearth$ of heavy elements \citep{BaraffePPVI}. This situation can be of course solved, at least for Jupiter, by increasing the accretion rate of solids  (Sect. \ref{other_Mzdot}). However, for intermediate mass planets, an increase of the solid accretion rate would make formation times so short, that they would likely become gas giants. So the combination of low opacities and high accretion rates of solids could work for forming Jupiters, but  forming Neptunes would be more challenging. 

We note that the dust opacities from \citet{Mordasini14} are analytical, and thus, many simplifications were considered to compute them. Perhaps the most important one is the assumption of a predominant grain size. \citet{Mordasini14} mention that if planetesimal ablation is important in all layers of the envelope, then the constant supply of small grains could raise the dust opacities. It is likely that in our enriched scenario this plays a relevant role.

\subsection{Exoplanets}
Our formation model assumes that envelopes get enriched in water during formation, because icy planetesimals are expected to be water rich. 
Water has indeed been detected in all the atmospheres of the giant planets of the solar system, although in very small amounts, even less than what is expected based on the detection of other volatile molecules, such as CH$_4$ \citep{galileo_probe, Irwin14}. The problem with measuring water on the outermost layers of the giant planets is that due to the low temperatures, this species is expected to have condensed deeper in, and therefore be present in the form of clouds. 

In this sense, transiting exoplanets offer a better opportunity to detect water in their atmospheres: since the equilibrium temperatures are much higher (these planets are much closer to their central star than the giants of the solar system, due to observational bias), water is expected to be present in the form of vapour. Indeed, of the 19 transiting planets whose spectra has been measured, 10 of them present signatures of water vapour in their atmospheres \citep[the remaining 9 are thought to posses water as well, but their signature to be obscured by clouds,][]{Iyer15}. From these 10 exoplanets, 9 have masses in the range of 0.5 - 2 jovian masses, so of course, these planets are expected to be hydrogen-helium rich. In other words, the amount of water present is expected to be low, as our results suggest (envelope metallicity should decrease with increasing planet mass). 

\begin{figure}
\begin{center}
\includegraphics[width=\columnwidth]{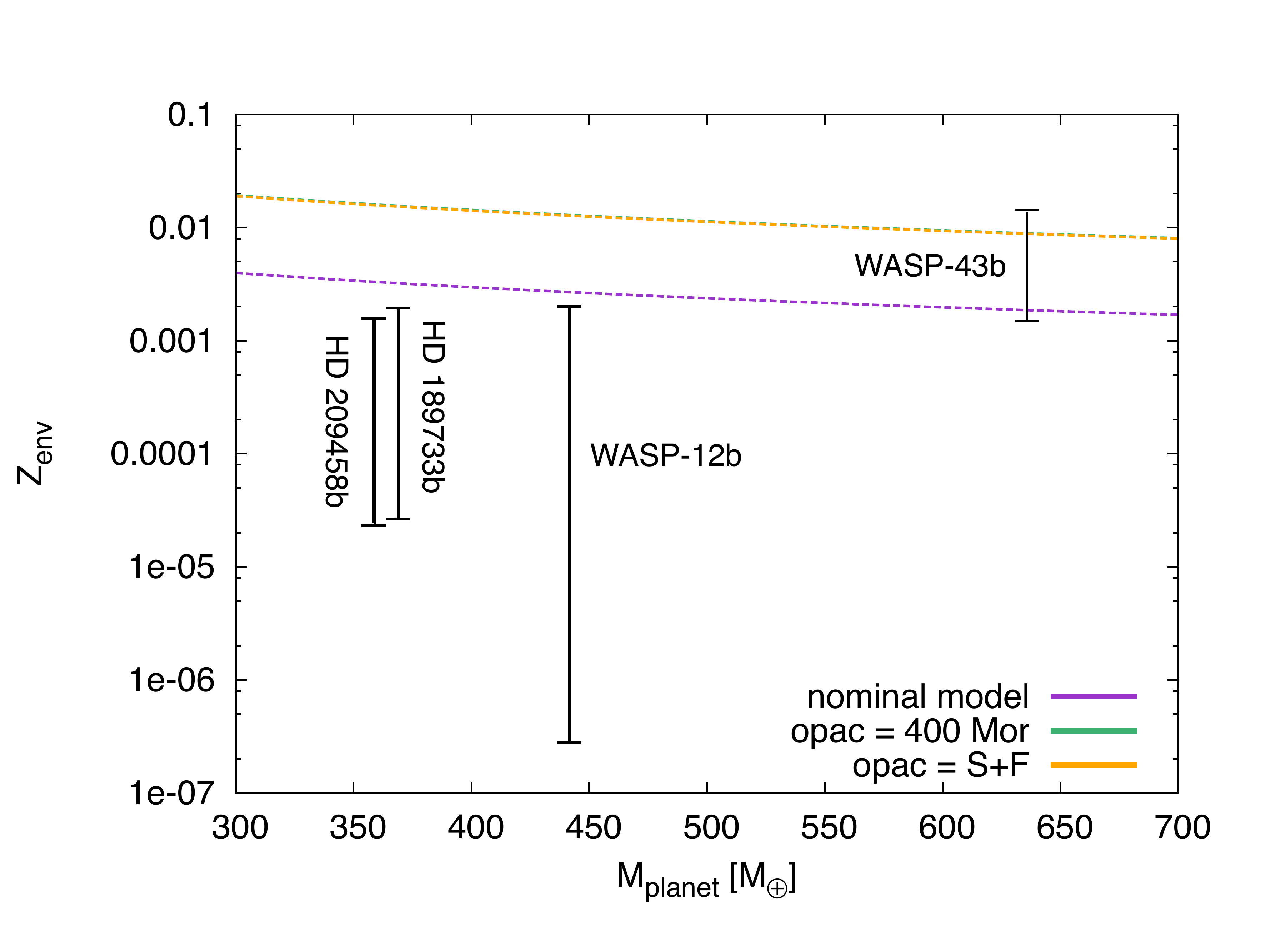} 
   \caption{Same as Fig. \ref{Z_Mp}, but we show just the mass range corresponding to four hot Jupiters where estimations of atmospheric mass fraction of water ($\Zenv$) have been reported, and we overplot these estimations.}
    \label{hot_jup}
\end{center}
\end{figure}

Two works report precise water abundances of hot Jupiter's atmospheres:  \citet{Kreidberg14}  for WASP-43b and \citet{Madhu14} for  HD 189733b, HD 209458b and WASP-12b. We have converted the abundances they give of water mixing ratios into mass fraction of water ($\Zenv$)\footnote{See explanation for this conversion in Appendix \ref{mix_ratio_to_Zenv}.} and plotted the predictions of our models together with these estimations in Fig. \ref{hot_jup}. We note that in the case of WASP-43 b, the predictions of both high and low opacity models work surprisingly well to explain the estimated abundance of water vapour. For the hot Jupiters reported by  \citet{Madhu14}, we find that all our models predict a larger mass fraction of atmospheric water, at least by a factor of 2. \citet{Madhu14} claim as well that the water abundances they find seem too low, and suggest that the presence of clouds might be obscuring some of the molecular features of the spectra \citep[hypothesis supported as well by][]{Sing16, Benneke15}. \citet{Madhu14} suggest, alternatively, that the atmospheres of these hot-Jupiters could bear more carbon-rich species than oxygen-rich. It is important to remark that in our formation model, we assumed that all the volatiles deposited in the planet's envelope  were just water (because of the equation of state). In this sense, it could perfectly be that we overestimate the amount of water, since our model neglects the possibility of initial amounts of, e.g, CH$_4$, CO$_2$, CO.

Another interesting case to link our results with observations is the detection of water in HAT-P-11b \citep{Fraine14}, the only Neptune-mass exoplanet where the presence of water vapor has been inferred from transmission spectroscopy. \citet{Fraine14} report from retrieval models that the atmosphere of HAT-P-11b is expected to have a metallicity of 1 to 700 times solar, with no real preference for a specific value within this range (Benneke, priv. comm.). Unfortunately, this wide range could imply an envelope metallicity from solar up to $\Zenv \sim  0.9$. HAT-P-11b has a total mass of $\Mp \approx 26$ $\Mearth$, so our models would predict a $\Zenv$ in the range of, approximately, 0.05 - 0.4 (see Fig. \ref{Z_Mp}); which would correspond to values of ``metallicity times solar" between 1 and 100. More precise measurements would be needed in order to determine if our models are able to predict the metal atmospheric content of this planet or not. 

A last word regarding the detection of water in exoplanets: it is important to remark that even if the sample of transmission spectra measured is still small, water shows to be common in the atmospheres of planets. This reinforces a formation scenario beyond the iceline, with volatile-rich planetesimals being dissolved in the atmospheres during formation. Even if in the future it is found that statistically the atmospheres are water poor, this could be a hint of an initial planetesimal composition that is more carbon rich \citep{Madhu14}. This is why it is important to determine precisely other volatile abundances besides water. 

\section{Conclusions}\label{conclusions}
We have performed the first self-consistent calculation of the growth of a planet including the effect of envelope enrichment due to the dissolution of icy planetesimals/pebbles. We have implemented suited equations of state and opacities taking into account different metallicities. Moreover, we have considered two different sets of opacities in order to test the impact on our results, which can be summarised as:

\begin{itemize}
\item Envelope enrichment accelerates notably the formation of gas giants. This is mainly a consequence of the increase of the mean molecular weight of the envelope. The thinner the envelope (i.e, the smaller the planetesimal), the sooner envelope enrichment sets in and the shorter the timescale to form a giant planet.

\item When envelope enrichment is taken into account, low and intermediate mass planets (namely mini-Neptunes to Neptunes) can be formed, with total mass fractions of H-He up to $30 \%$, this number being independent of the choice of opacities.

\item Low-opacities allow for the formation of mini-Neptunes, whereas high-opacities lead to the formation of Neptunes. 

\item High-opacities are preferable for explaining the total mass and metallicity of the giant planets of the solar system. 

\item We were able to quantify the amount of volatile material remaining in the primordial atmospheres as a result of formation. These allowed us to compare our results with water abundances inferred from the transmission spectra of transiting exoplanets. We find good agreement for WASP-43b, whereas for the other three cases, the measured water abundance is lower than what is predicted by our models. 

\item We predict that envelope metallicity and total metallicity should decrease with planetary mass. 
\end{itemize}

\paragraph{\textit{Acknowledgments}.}We thank M. Ikoma for fruitful discussions and F. Benitez for providing the EOS tables and revising this manuscript. J.V. acknowledges financial support from the Swiss-based MERAC Foundation. This work has, in part, been carried out within the framework of the National Centre for Competence in Research PlanetS supported by the Swiss National Science Foundation. The authors acknowledge financial support from the SNSF.

\begin{appendix}
\section{Conversion of water mixing ratios to water mass fractions}\label{mix_ratio_to_Zenv}
It is a common practice in papers that estimate abundance of different compounds to give them in terms of ``volume mixing ratios". In particular, for estimates on water abundances derived from transmission spectroscopy measurements, the volume mixing ratio of water ($m_r$) is defined as \citep{Madhu14}:

\begin{equation}
m_r = \frac{n(\text{H$_2$O})}{n(\text{H}_2)} ,
\end{equation} 
where $n$(H$_2$O) and $n$(H$_2$) are the number density of water molecules and hydrogen, respectively.

Because of the molar masses of H$_2$O and H$_2$, in terms of mass, the ratio of water to hydrogen reads:
\begin{equation}
\frac{m_{\text{H$_2$O}} }{m_{\text{H}_2}} =  \frac{18}{2}\frac{n(H_2O)}{n(H_2)} = 9 m_r .
\end{equation}
 
 In the formation models presented here, we defined the envelope metallicity as the mass fraction of water in the envelope. Thus,
 \begin{align}\label{mr_to_Zenv}
 \Zenv =& \frac{m_{\text{H$_2$O}}}{m_{\text{H}_2} + m_{\text{He}} + m_{\text{H$_2$O}}} \nonumber \\
           =& \frac{m_{\text{H$_2$O}} / m_{\text{H}_2}}{ 1 + Y/X +  m_{\text{H$_2$O}} / m_{\text{H}_2}} \nonumber \\
           =& \frac{9 m_r}{1 + (Y/X)_{\odot} + 9 m_r} ,
 \end{align} 
where we assumed the ratio of helium to hydrogen to be solar. 
This last equation is the one we use to estimate the values of $\Zenv$ shown in Fig. \ref{hot_jup} for the different hot-Jupiters whose water mixing ratios has been determined \citep{Madhu14, Kreidberg14}.

\end{appendix}

\bibliographystyle{aa}
\bibliography{lit}{}

\end{document}